\documentclass[pdflatex,sn-mathphys-num]{sn-jnl}
\usepackage{arabtex}
\usepackage{utf8}
\setcode{utf8}
\usepackage[utf8]{inputenc} 

\newcommand{\mypersian}[1]{%
  {\fontsize{9pt}{10pt}\selectfont\rmfamily\RL{#1}}%
}

\usepackage{mathrsfs}
\usepackage{graphicx}%
\usepackage{multirow}%
\usepackage{amsmath,amssymb,amsfonts}%
\usepackage{amsthm}%
\usepackage{mathrsfs}%
\usepackage[title]{appendix}%
\usepackage{xcolor}%
\usepackage{textcomp}%
\usepackage{manyfoot}%
\usepackage{booktabs}%
\usepackage{algorithm}%
\usepackage{algorithmicx}%
\usepackage{algpseudocode}%
\usepackage{listings}%
\usepackage{subcaption}%
\usepackage{placeins}%
\usepackage{diagbox}%

\theoremstyle{thmstyleone}%

\theoremstyle{thmstyletwo}%

\theoremstyle{thmstylethree}%

\begin{document}

\title[Article Title]{PSRB: A Comprehensive Benchmark for Evaluating Persian ASR Systems}

\author[1]{\fnm{Nima} \sur{Sedghiyeh}}
\author[1]{\fnm{Sara} \sur{Sadeghi}}
\author[1]{\fnm{Reza} \sur{Khodadadi}}
\author[1]{\fnm{Farzin} \sur{Kashani}}
\author[1]{\fnm{Omid} \sur{Aghdaei}}
\author[1]{\fnm{Somayeh} \sur{Rahimi}}
\author[1]{\fnm{Mohammad Sadegh} \sur{Safari}}

\email{\{nima.sedghiyeh, sara.sadeghi, reza.khodadadi, farzin.kashani, omid.aghdaei, somayeh.rahimi, mohammadsadeq.safari\}@partdp.ai}

\affil*[1]{\orgdiv{Part AI Research Center}, \state{Tehran}, \country{Iran}}


\abstract{Although Automatic Speech Recognition (ASR) systems have become an integral part of modern technology, their evaluation remains challenging, particularly for low-resource languages such as Persian. This paper introduces Persian Speech Recognition Benchmark(PSRB), a comprehensive benchmark designed to address this gap by incorporating diverse linguistic and acoustic conditions. We evaluate ten ASR systems, including state-of-the-art commercial and open-source models, to examine performance variations and inherent biases. Additionally, we conduct an in-depth analysis of Persian ASR transcriptions, identifying key error types and proposing a novel metric that weights substitution errors. This metric enhances evaluation robustness by reducing the impact of minor and partial errors, thereby improving the precision of performance assessment. Our findings indicate that while ASR models generally perform well on standard Persian, they struggle with regional accents, children’s speech, and specific linguistic challenges. These results highlight the necessity of fine-tuning and incorporating diverse, representative training datasets to mitigate biases and enhance overall ASR performance. PSRB provides a valuable resource for advancing ASR research in Persian and serves as a framework for developing benchmarks in other low-resource languages.
A subset of the PSRB dataset is publicly available\footnote{https://huggingface.co/datasets/PartAI/PSRB}.}


\maketitle

\section{Introduction}\label{sec1}

Automatic Speech Recognition (ASR) systems have become deeply integrated into modern life, powering virtual assistants, automated subtitles, and customer service applications. As their role expands, thorough evaluation is essential to ensure reliability across diverse real-world conditions. While state-of-the-art (SOTA) ASR models \cite{rekesh2023fast,chung2021w2v,radford2023robust} achieve impressive accuracy on benchmarks like LibriSpeech \cite{panayotov2015librispeech}, these evaluations often overestimate real-world performance due to controlled conditions and structured speech datasets \cite{aksenova2021might,szymanski2020we}. Spontaneous speech, diverse speaker demographics, and noisy environments present significant challenges that many ASR benchmarks fail to capture. These limitations highlight the need for a more comprehensive benchmark that evaluates ASR models across diverse and challenging speech environments.

This gap between benchmark results and real-world applicability highlights two key limitations. First, ASR systems suffer from domain mismatch, where models trained on controlled data struggle with unrepresented contexts such as domain-specific terms or spontaneous speech. Second, demographic biases in ASR systems disproportionately affect underrepresented groups, such as children, pre-teen speakers, and non-standard accent speakers. These challenges are particularly acute for low-resource languages like  Persian, where limited training data and evaluation resources exacerbate performance disparities.

Spoken by over 100 million people, Persian poses unique ASR challenges due to its linguistic diversity—spanning regional accents like Baluchi and Kurdish—and features like variable word boundaries. Yet, comprehensive benchmarks for Persian ASR remain scarce. This study addresses these gaps by introducing the Persian Speech Recognition Benchmark (PSRB), a novel benchmark designed to assess ASR performance across varied linguistic, demographic, and acoustic conditions. We evaluate ten state-of-the-art commercial and open-source ASR models, analyze their error patterns, and propose a new metric—Substitution Weighted WER (SW-WER)—to enhance evaluation precision. Our findings reveal critical weaknesses in current systems, particularly with informal speech and regional accents, and offer insights for developing more robust and inclusive Persian ASR technologies. 

The rest of the paper is structured as follows. Section \ref{sec2} reviews related work on ASR benchmarking, highlighting the limitations of existing datasets. Section \ref{sec3} introduces the proposed framework, detailing PSRB and the evaluation metrics, including the newly introduced robust metric for evaluations. Section \ref{sec4} presents the benchmark results, comparing the performance of various ASR models, including open-source and commercial systems. Section \ref{sec5} conducts an in-depth error analysis, categorizing the most common transcription mistakes in Persian ASR. Section \ref{sec6}  evaluates robustness, examining model performance variations based on formality, noise levels, speaker demographics, and spontaneity. Finally, Section \ref{sec7}  concludes the study, summarizing key findings and outlining directions for future research in Persian ASR benchmarking.

\section{Related Work}\label{sec2}

A critical aspect of developing an ASR system is evaluating its performance on diverse, previously unseen data to identify errors, diagnose issues, and derive valuable insights. Although ASR models are typically trained on standardized, clean, or read speech data, they often struggle in complex real-world scenarios, such as noisy environments or multi-speaker interactions. This challenge, commonly referred to as domain mismatch, arises from the disparity between training conditions and real-world deployment settings. Consequently, the development of ASR benchmarks has emerged as a key research area, with increasing recognition of their limitations in capturing the full spectrum of real-world speech variability.

Traditional benchmarks, such as LibriSpeech \cite{panayotov2015librispeech}, primarily focus on clean, well-structured speech, rendering them inadequate for assessing ASR performance in spontaneous or conversational settings. Studies like those by Rio et al. \cite{del2021earnings} and Cao et al. \cite{cao2023comparative} underscore the need for benchmarks that evaluate ASR systems under such challenging conditions. Specifically, Rio et al. \cite{del2021earnings} propose benchmarks tailored for "ASR in the wild," while Cao et al. \cite{cao2023comparative} reveal significant performance gaps in noisy, multi-speaker classroom environments. These findings highlight the importance of incorporating spontaneous and multi-domain speech into benchmarks.

To mitigate domain mismatch and improve cross-domain generalization, the End-to-End Speech Benchmark (ESB) \cite{gandhi2022esb} was introduced. ESB evaluates ASR models across diverse domains and datasets without prior knowledge of data distributions, promoting a multi-domain assessment. However, although ESB accounts for varied acoustic environments and dataset diversity, it overlooks critical factors such as data sources, speaker age, gender, and other demographic characteristics. As Aksënova et al. \cite{aksenova2021might} emphasize, comprehensive benchmarks should encompass diverse linguistic contexts—including variations in speech styles, acoustic conditions, and speaker demographics—to ensure a thorough and robust evaluation of ASR systems. This need has spurred increased research efforts to develop benchmarks that accurately reflect real-world conditions and assess model performance beyond conventional metrics like Word Error Rate (WER). Similarly, Szymański et al. \cite{szymanski2020we} Kuhn et al. [11] note that standard benchmarks often fail to account for WER variability across complex and challenging domains. 

While most benchmark studies focus on English ASR, challenges persist in low-resource languages like Persian, where dataset limitations further exacerbate model biases. Existing Persian ASR datasets provide a partial foundation but have significant shortcomings. The Persian language poses unique challenges for ASR benchmarking due to its linguistic diversity and regional accents. Existing resources, such as the DeepMine dataset by Zeinali et al. \cite{zeinali2019multi}, provide a robust foundation, featuring over 1,850 speakers and approximately 480 hours of audio recordings across a diverse demographic range. The test set of the DeepMine dataset comprises approximately six hours of audio. However, its focus on formal speech limits its applicability to informal or spontaneous contexts. Similarly, the Persian subset of Mozilla Common Voice \cite{ardila2019common} offers valuable data for multilingual speech research but suffers from demographic imbalances and data quality concerns due to its crowdsourcing approach. The FLEURS dataset \cite{conneau2023fleurs}, while useful for few-shot learning, focuses on short, controlled utterances, which are insufficient for evaluating conversational or informal speech.

Beyond dataset limitations, demographic biases in ASR performance remain a significant concern. For instance, Fuckner et al. \cite{fuckner2023uncovering} identify systemic disparities in WER for non-native speakers, children, and elderly users when using SOTA models such as Wav2Vec2 and Whisper. These performance gaps are attributed to imbalanced training datasets that inadequately represent key demographic groups, including regional accents and age variations. Likewise, Feng et al. \cite{feng2024towards} stress the importance of inclusivity in ASR systems, advocating for fairness metrics and diverse datasets to improve accuracy for underrepresented populations. Similarly, Schubert et al. \cite{schubert2024challenges} highlight comparable challenges in German ASR, particularly in recognizing multi-ethnolectal speech among adolescents, while Kulkarni et al. \cite{kulkarni2024balancing} propose solutions for reducing bias in Portuguese ASR systems through balanced data augmentation. 

Effective ASR evaluation extends beyond WER analysis to detailed error diagnostics. After benchmarking an ASR model, conducting error analysis helps uncover recurring patterns in misrecognitions, guiding improvements. By systematically analyzing error patterns, potential solutions can be devised to enhance model performance. Error analysis frameworks developed for other languages can inform improvements in Persian ASR systems. For example, Wirth and Peinl \cite{wirth2022asr} provide a detailed error analysis for German ASR, while Schubert et al. \cite{schubert2024challenges} address multi-ethnolectal challenges among German adolescents. These studies emphasize the role of linguistic diversity in error analysis and offer strategies for mitigating biases that could be adapted to the Persian context. Given the limitations of Persian language datasets and ASR systems, as well as the importance of accurate evaluation, we propose a benchmark for conducting error analysis and assessing model robustness. This benchmark provides researchers and engineers in Persian ASR with a structured framework to identify key challenges, refine models, and implement solutions to enhance speech recognition performance in Persian.

\section{Proposed Framework}\label{sec3}

A robust ASR evaluation requires two fundamental components: a comprehensive benchmark that represents diverse linguistic and acoustic conditions and a well-designed evaluation metric that accurately measures system performance. This section presents the PSRB and the evaluation metrics used in this study. Section \ref{subsec3.1} describes the benchmark, covering the data collection process (Section \ref{subsubsec3.1.1}), which includes defining benchmark criteria, identifying and collecting data sources, data processing, and annotation and quality control. It also provides statistical insights into the dataset (Section \ref{subsubsec3.1.2}). Section \ref{subsec3.2} focuses on evaluation metrics, introducing a novel metric that combines WER and CER to enhance performance assessment.

\subsection{Benchmark}\label{subsec3.1}

\subsubsection{Data collection process}\label{subsubsec3.1.1}

\paragraph{Defining Benchmark Criteria}\label{para1}

\begin{figure*}[htbp]
  \centering
  \includegraphics[height=6.5cm, width=\linewidth]{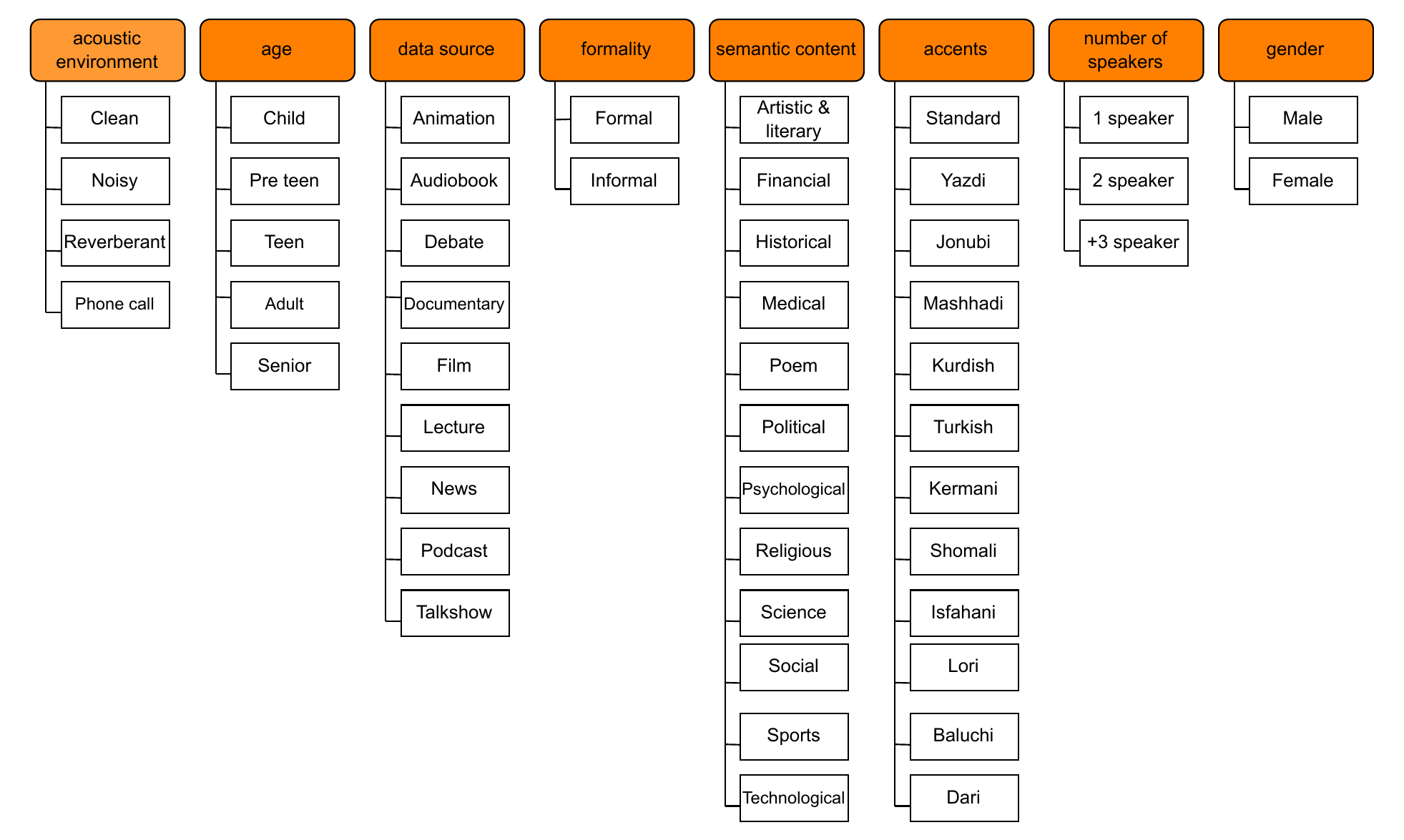}
  \vspace{-2mm}
  \caption{Overall structure of the PSRB benchmark. Diagram illustrating the multidimensional structure of the PSRB benchmark, encompassing age, gender, accent, data source, spontaneity, semantic content, acoustic environment, and formality to evaluate Persian ASR systems comprehensively.}
  \label{fig:metadata}
  \vspace{-5mm}
\end{figure*}

Through a comprehensive analysis of key factors influencing the development of a reliable ASR benchmark, we identified eight categories, as illustrated in Figure \ref{fig:metadata}. Our primary objective was to create a well-balanced benchmark that accurately represents real-world scenarios. These categories include semantic content, data sources, speech style, linguistic diversity, speaker diversity, accent variation, multi-speaker scenarios, and acoustic environments. Each category was carefully structured to encompass the diverse characteristics of Persian speech, ensuring broad coverage and a fair evaluation of ASR performance.

\paragraph{Identifying and Collecting Data Sources}\label{para2}

In the next phase, we searched for Persian resources that met our predefined standards, ensuring comprehensive coverage across all categories. Our dataset is built from a wide range of sources, including podcasts, news archives, and video streaming platforms. Sources such as Telewebion\footnote{https://telewebion.com}, Aparat\footnote{https://www.aparat.com}, YouTube\footnote{https://www.youtube.com}, and Iranseda\footnote{https://iranseda.ir} offered diverse movies, TV series, talk shows, lectures, and news broadcasts. 

\paragraph{Data Processing}\label{para3}

All collected files were converted to a standard audio format with a consistent sample rate to ensure integrity. The audio files were then segmented into clips ranging from a minimum of one second to a maximum of 100 seconds to optimize the benchmark usability. Clips were carefully selected from these segments based on a predefined structure to ensure a well-balanced benchmark that adequately represents all categories.

\paragraph{Annotation and Quality Control}\label{para4}

Generating high-quality transcriptions for each audio segment was a fundamental aspect of our process. The first step involved an in-depth linguistic analysis of Persian grammar conducted by a team of expert data scientists, leading to the development of a unified annotation guideline. A small portion of the dataset was initially labeled based on these guidelines.

This labeled subset served as the basis for quality assessment, where multiple annotators' performances were evaluated. From this evaluation, four top-performing annotators were selected for their accuracy and consistency.

The annotation process followed a two-tier review system:
 First, the most precise annotators performed an initial review, followed by a final verification by the technical team to ensure linguistic accuracy and compliance with the established guidelines. Throughout this process, two key principles guided the work: (1) Consistency, ensuring uniform transcription and adherence to Persian grammar rules, and (2) Standardization, maintaining a structured approach to ensure the benchmark's reliability.

After collecting the speech dataset and completing the metadata using category labels and the corresponding text for each audio file, the final dataset is prepared. An example of this is presented in Table \ref{tab:metadata_examples}.

\begin{table*}[h]
    \centering
    \renewcommand{\arraystretch}{1.2}
    \resizebox{\textwidth}{!}{
    \begin{tabular}{l p{5cm} p{5cm}}
        \toprule
        \textbf{Attribute} & \textbf{Entry 1} & \textbf{Entry 2} \\
        \midrule
        \textbf{Audio} & 1368-f370.wav & 3123-p641.wav \\
        \hline
         \textbf{Text} & {\small \RL{
 مو نمی‌خواستُم مسخره‌بازی دربيارُم مو نمی‌دونستُم کيه اين مرده.
        }}
        & {\small \RL{
            کامپيوترها به کمک هوش مصنوعی،  }}
           {\small \RL{ عملکرد بهتری در شنيدن، خواندن،  }}
         {\small \RL{   تماشا و درک کردن پيدا می‌کنند.
        }} \\
        \hline
        \textbf{Number of Speakers} & 2 & 1 \\
        \hline
        \textbf{Gender} & Male & Female \\
        \hline
        \textbf{Age} & Mix & Adult \\
        \hline
        \textbf{Accents} & Jonubi & Standard \\
        \hline
        \textbf{Formality} & Informal & Formal \\
        \hline
        \textbf{Semantic Content} & Social & Technological \\
        \hline
        \textbf{Data Source} & Film & Podcast \\
        \hline
        \textbf{Acoustic Environment} & Clean & Noisy \\
        \hline
        \textbf{Spontaneous} & No & No \\
        \bottomrule
    \end{tabular}
    }
    \caption{Two examples from benchmark metadata}
    \label{tab:metadata_examples}
    \vspace{-5mm}
\end{table*}

\subsubsection{Statistics}\label{subsubsec3.1.2}

\begin{figure*}[htbp]
  \centering
  \includegraphics[height=4.6cm, keepaspectratio]{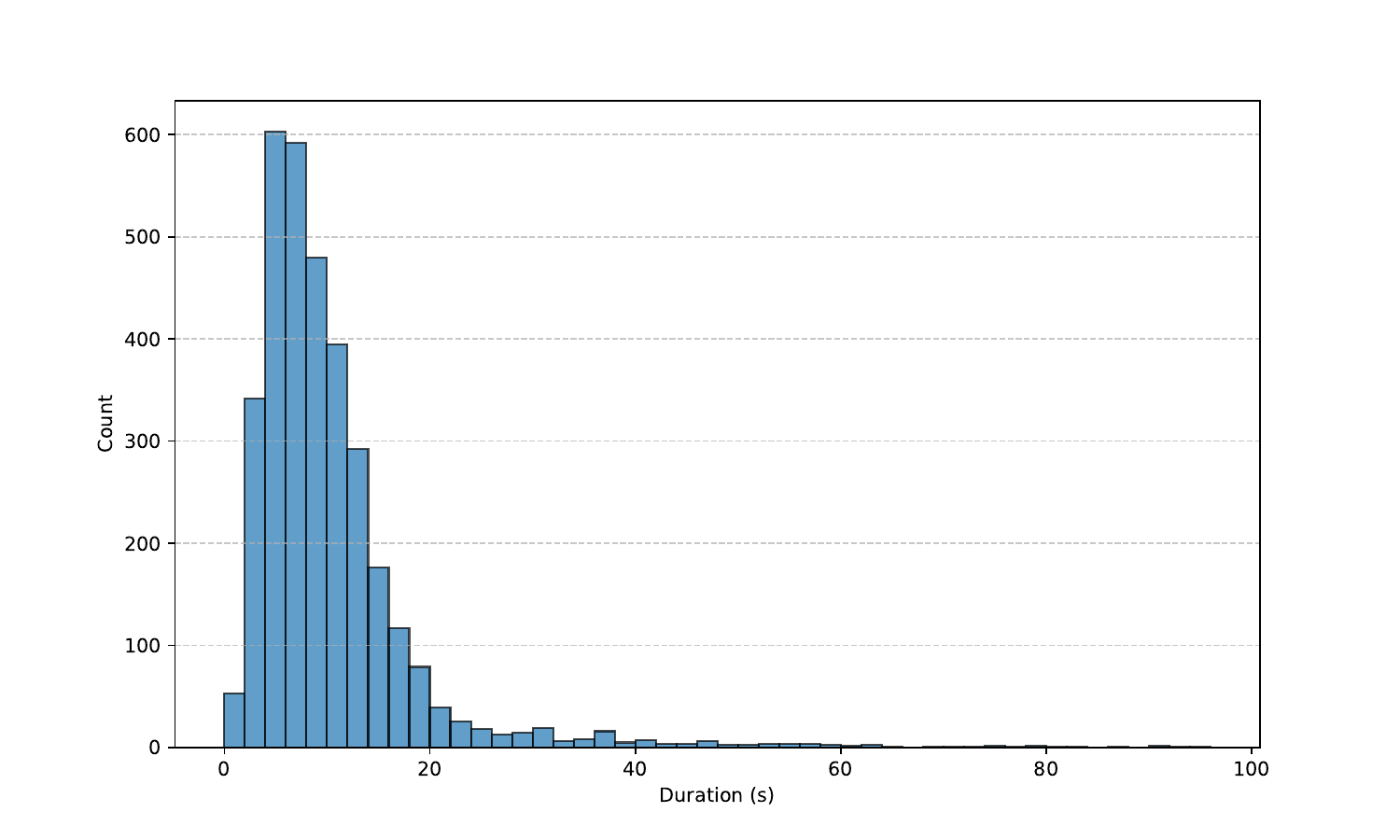}
  \vspace{-1mm}
  \caption{Histogram of benchmark audios duration }
  \label{fig:duration_hist}
  \vspace{-5mm}
\end{figure*}

As mentioned in the previous subsection, the collected data was segmented into smaller audio files. This approach facilitated cleaner and more accurate transcription by annotators. Large audio files were excluded to better assess ASR precision in real-world applications. The focus of this evaluation is on the models' accuracy rather than their response time or computational efficiency. The benchmark enforces audio duration constraints between 1 and 100 seconds for consistency. As observed in the corresponding plot, the majority of the audio data falls within the 4 to 10-second range, with relatively few samples exceeding 40 seconds. The distribution of audio durations is illustrated in the histogram shown in Figure \ref{fig:duration_hist}.

\vspace{-5mm}

To comprehensively represent real-world conditions, our dataset includes a substantial portion of spontaneous speech (58.6\%) alongside structured, read-speech recordings. It covers a range of acoustic environments, with 47.33\% clean speech and 52.67\% encompassing noisy, phone call, and reverberant conditions. Furthermore, the dataset maintains a balance in formality, comprising 28.6\% formal and 71.3\% informal speech, effectively capturing the linguistic diversity found in everyday communication.

Table \ref{table:datasets_comparison} compares existing Persian test sets with our benchmark, which provides sufficient duration for comprehensive evaluation. Additionally, our dataset includes longer speech samples compared to DeepMine ASR test-set, allowing us to assess ASR models on medium-length inputs with diverse semantic content. The average duration of our samples is 11 seconds, whereas the CommonVoice test set has a significantly lower average, primarily covering short utterances with limited semantic depth.

\begin{table*}[h]
    \centering
    \renewcommand{\arraystretch}{1.2}
    \setlength{\tabcolsep}{4pt} 
    \small
    \vspace{-1mm}
    \resizebox{\textwidth}{!}{
    \begin{tabular}{lcccccccc}
        \toprule
        \textbf{Benchmark} & \textbf{Dur.(h)} & \textbf{Min/Max Dur.(s)} & \textbf{Avg Dur.(s)} & \textbf{\#Utts} & \textbf{\#Wrds} & \textbf{\#Unq.Wrds} & \textbf{SR(Hz)} & \textbf{\#Spks} \\
        \midrule
        Fleurs   & 3.7   & 5.3 / 39.2   & 15.3  & 870   & 22789  & 2585   & 16k  & 324  \\
        CV      & 14.2  & 0.07 / 105.8  & 4.9  & 10404 & 53180  & 13279  & 32k,48k  & 1741 \\
        Deepmine & 6.5   & 1.8 / 25.04    & 9.2  & 2526  & 43966  & 7774   & 16k  & 50  \\
        PSRB& 10.4  & 0.8 / 383.7  & 11.0  & 3372  & 78526  & 12461  & 16k  & 756  \\
        \bottomrule
    \end{tabular}
    }
    \caption{Comparison of various Persian ASR datasets based on key characteristics, including duration, average duration, total word count, unique word count, number of utterances, and speaker diversity.}
    \label{table:datasets_comparison}
    \vspace{-5mm}
\end{table*}

Our dataset ensures reliability and accuracy by including at least 50 utterances per category, creating a balanced and representative benchmark for ASR evaluation. While CommonVoice has the highest number of utterances, its crowd-sourced nature introduces inconsistencies in pronunciation and transcription, making it less suitable for complex linguistic analysis. Our benchmark consists of 78,526 words, including 12,461 unique words from diverse topics, offering broader lexical coverage. In contrast, CommonVoice, despite having more unique words, is primarily focused on social content. Additionally, our dataset includes a more diverse range of speakers across different age and gender groups, addressing the demographic limitations of CommonVoice for ASR evaluation.

\subsection{Metrics}\label{subsec3.2}

To assess the performance of Persian ASR systems, CER is used as the primary evaluation metric. Compared to WER, CER is more suitable for Persian due to its unique linguistic complexities, including orthographic variations, diverse word formations, and word boundary ambiguities. CER provides a finer assessment of transcription accuracy, making it effective for Persian ASR systems. However, a limitation of CER is that it operates at the character level without considering linguistic structure or the semantic validity of generated words.

Therefore, we introduce a new metric called Substitution Weighted WER (SW-WER), which is based on error rate and combines WER and CER, providing valuable insights into ASR model performance. This metric is defined as follows:

\begin{equation}
\text{SW-WER} = \frac{S + I + D}{N_{\text{sub}} + C + D}
\end{equation}
\vspace{-1mm}

where:
\( C \) is the number of correct (hit) words, \( I \) is the number of inserted words, \( D \) is the number of deleted words, and \( S \) is computed from substitution errors.

The substitution error \( S \) is computed as:

\begin{equation}
S = \sum_{i=1}^{N} s_i
\end{equation}

where each substitution error \( s_i \) is defined as:

\begin{equation}
s_i = n_i \cdot \text{CER}(\text{str}_1, \text{str}_2)
\end{equation}

where:
\( n_i \) is the number of words in the reference substitution segment, and \( \text{CER}(\text{str}_1, \text{str}_2) \) is the Character Error Rate (CER) computed between the reference and hypothesis segments. The CER is bounded by:

\begin{equation}
0 \leq \text{CER}(\text{str}_1, \text{str}_2) \leq 1
\end{equation}

Additionally, the total number of substituted words across all substitution segments is denoted as:

\begin{equation}
N_{\text{sub}} = \sum_{i=1}^{N} n_i
\end{equation}
\vspace{-1mm}

SW-WER modifies the traditional WER calculation by addressing a key limitation. When aligning two Persian texts using Levenshtein distance, some substitutions involve only minor character changes. As ASR increasingly serves as the front-end for LLMs to process input speech, LLMs can detect and correct ASR errors using their linguistic understanding \cite{ma2024asr}. While minor character-level mistakes have little impact, incorrect word generation can severely affect Natural Language Processing (NLP) tasks such as summarization and sentiment analysis. Yet, the standard WER formula treats each substitution as a full-word error. Instead, the CER is computed for each substitution and multiplied by the number of words in that alignment.

\begin{figure*}[htbp]
  \centering
  \includegraphics[height=4cm, width=0.9\linewidth]{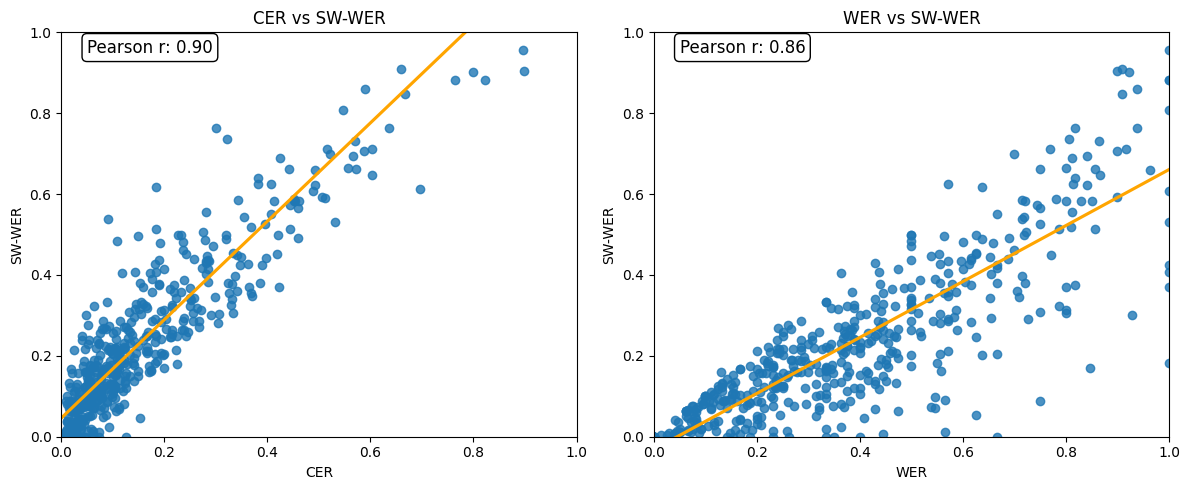}
  \vspace{-2mm}
  \caption{Scatter plots demonstrating the correlation between WER, CER, and the proposed SW-WER metric.}
  \label{fig:metricscorelation}
  \vspace{-5mm}
\end{figure*}

To analyze the relationship between the newly proposed metric and established metrics such as WER and CER, the following plots depict its measurement on the introduced benchmark and its correlation with these existing metrics. To enhance visualization, 500 data points were randomly sampled and plotted in Figure \ref{fig:metricscorelation}. As observed in the plots, the strong correlation indicates that this metric can be effectively utilized for future analysis of model performance and error assessment.

\begin{table}[htbp]
    \centering
    \renewcommand{\arraystretch}{1}
    \resizebox{\textwidth}{!}{
    \begin{tabular}{|l|c|c|c|}
        \hline
        \textbf{Example} & \textbf{CER} & \textbf{WER} & \textbf{SW-WER} \\
        \hline
        \multirow{2}{*}{Ref: \mypersian{  به نام خداوند بی‌نظير و بی‌همتا
        }} & \multirow{2}{*}{12.90} & \multirow{2}{*}{100.00} & \multirow{2}{*}{33.33} \\
        & & & \\
        Hyp: \mypersian{ بنام خداوند بی نظير و بی همتا } & & & \\
        \hline
        \multirow{2}{*}{Ref: \mypersian{  به جرأت ميتوان گفت برجسته‌ترين نخبگان ايران آن‌جا گرد هم آمدند }} & \multirow{2}{*}{12.90} & \multirow{2}{*}{81.81} & \multirow{2}{*}{43.98} \\
        & & & \\
        Hyp: \mypersian{به جرئت می توان گفت برجسته ترين نخبگان ايران آن جا گردهم آمده‌اند } & & & \\
        \hline
        \multirow{2}{*}{Ref: \mypersian{ کلام منثور پيچيده‌ تر از وزن‌ ها و بحرهای شعر کلاسيک است }} & \multirow{2}{*}{8.92} & \multirow{2}{*}{50.00} & \multirow{2}{*}{28.15} \\
        & & & \\
        Hyp: \mypersian{ کلام منصور پيچيده‌تر از وزنه‌ها و بهرهای شعر کلاسيک است } & & & \\
        \hline
        \multirow{2}{*}{Ref: \mypersian{ بچه‌ها در حياط مدرسه فوتبال بازی می‌کردند }} & \multirow{2}{*}{9.52} & \multirow{2}{*}{42.85} & \multirow{2}{*}{26.03} \\
        & & & \\
        Hyp: \mypersian{ بچه‌ها در حياط مدل سه فوتبال بازی ميکردند } & & & \\
        \hline
    \end{tabular}
    }
    
    \caption{ Examples of hypothesis with SW-WER. Metrics are reported as percentages}
    \label{table:error_rates}
    \vspace{-5mm}
\end{table}

SW-WER provides a more balanced evaluation of ASR performance compared to CER and WER. Although CER focuses solely on character-level differences without considering language structure, making it an incomplete measure of transcription quality, WER relies on a rigid error classification. Additionally, WER is less effective for languages with complex morphology and flexible word structures, such as Persian, and tends to penalize short words more heavily, leading to inconsistencies across different languages and text types. In contrast, our proposed metric offers a more comprehensive assessment by integrating both character- and word-level errors. Table \ref{table:error_rates} compares these three metrics for ground truth and ASR output pairs. As shown in this table, the newly introduced metric provides a more accurate error evaluation, addressing cases where WER overestimates differences despite identical meanings (first two rows) and where CER underestimates differences despite semantic variation (last two rows).

\section{Benchmark Results}\label{sec4}

To assess the performance and robustness of ASR systems for the Persian language, we evaluate a diverse set of SOTA ASR models using our proposed benchmark. The evaluation focuses on analyzing model performance across a range of scenarios designed to mimic real-world conditions. The experiments included both open-source and commercial systems, covering a range of architectures and training methodologies. By analyzing accuracy and robustness under varied conditions, this evaluation provides a detailed understanding of the strengths and limitations of each model, offering valuable insights into their applicability for Persian ASR tasks.

For the Persian language, the following open-source models were evaluated: Whisper large-v3 (Whisper) \cite{radford2023robust} and Faster Whisper large-v3 (Faster-Whisper\footnote{https://github.com/SYSTRAN/faster-whisper}), fine-tuned version of Wav2vec 2.0 XLS-R \cite{babu2021xls} for Persian (SLPL W2V2\footnote{https://huggingface.co/SLPL/Sharif-wav2vec2}), Seamless M4T V2 large (Seamless) \cite{barrault2023seamless}, STT Fa FastConformer Hybrid Transducer-CTC Large (FC-Fa\footnote{https://huggingface.co/nvidia/stt\_fa\_fastconformer\_hybrid\_large}) \cite{gulati2020conformer, rekesh2023fast} from NVIDIA NeMo framework\footnote{https://github.com/NVIDIA/NeMo}, utilizing both Connectionist Temporal Classification (CTC) \cite{graves2006connectionist} and Transducer \cite{graves2012sequence} decoders, vosk-model-fa-0.5 (Vosk\footnote{https://alphacephei.com/vosk/models/vosk-model-fa-0.5.zip}) \cite{VoskSTT}, a DNN-HMM architecture based on Kaldi and Vosk library. Additionally, several commercial multilingual models, including Microsoft Azure (Azure) \cite{AzureSTT} and Google Chirp V2 (Chirp) \cite{GoogleSTT}, were assessed.
Among the commercial monolingual Persian ASR systems, the two best-performing models are included in the results: Avanegar\footnote{https://api.ivira.ai/partai/speechRecognition} and Aipa\footnote{https://aipaa.ir/demo/voice-analysis?tag=asr}.

For the evaluation of open-source models, Python scripts from each model's repository were utilized to automatically generate results, which were then analyzed by measuring CER, WER, and SW-WER for each utterance and calculating their averages. In contrast, for commercial models, their respective APIs were used to obtain transcriptions, as direct access to their models was not available. The results are presented in Table \ref{table:asr_performance_I}.

As shown in \ref{table:asr_performance_I}, among Persian commercial ASR models, Avanegar achieves the highest performance, with a CER of 8.75\% and a WER of 19.3\%. Aipa follows closely, with a CER of 10.43\%, performing slightly below Avanegar. Among multilingual commercial models, Chirp v2 demonstrates the best performance, achieving a CER of 9.05 and a WER of 19.92\%. In the open-source models, Faster-Whisper, a modified version of the Whisper architecture, achieves the highest accuracy while maintaining low latency. The Faster-Whisper model outperforms Whisper, as both share the same architecture; however, Faster-Whisper addresses many of the issues present in Whisper and exhibits significantly fewer hallucinations. This error is further discussed in Section \ref{subsec5.5}. However, other open-source models show minimal performance differences, producing comparable results.

The findings further indicate that commercial models outperform open-source models, primarily due to their access to large, high-quality, and diverse supervised datasets in Persian. In contrast, open-source models suffer from limited Persian-language data and labeling challenges. Additionally, models such as Seamless and Whisper struggle with Persian transcription despite their strong performance in English. This is because they rely on self-supervised learning, and the fine-tuning stage includes limited Persian data. Furthermore, these models are prone to hallucination errors, which will be discussed in the next section.

As previously discussed regarding SW-WER, this metric correlates strongly with WER and CER. The difference between WER and SW-WER indicates the impact of substitution errors on WER. For models such as Avanegar, Aipa, and Chirp, this difference is small, suggesting higher word prediction accuracy. In contrast, models like Faster-Whisper and SLPL W2V2 show larger discrepancies, reflecting lower prediction accuracy.

\begin{table}[htbp]
    \centering
    \renewcommand{\arraystretch}{1}
    \resizebox{0.5\textwidth}{!}{
    \begin{tabular}{lccc}
        \toprule
        \textbf{Model} & \textbf{CER} & \textbf{WER} & \textbf{SW-WER}\\
        \midrule
        Vosk & 23.96 & 44.62 & 39.41 \\
        Seamless & 22.30 & 38.85 & 34.76 \\
        Whisper & 18.92 & 41.49 & 36.97 \\
        Faster-Whisper & 13.72 & 33.93 & 26.77 \\
        SLPL W2V2 & 19.06 & 46.74 & 39.16 \\
        FC-Fa & 19.42 & 44.85 & 37.58 \\
        Azure & 15.78 & 33.94 & 28.06 \\
        Chirp & 9.05 & 19.92 & 15.99 \\
        Aipa & 10.43 & 24.64 & 19.83 \\
        Avanegar & \textbf{8.75} & \textbf{19.30} & \textbf{15.68} \\
        \bottomrule
    \end{tabular}
    }
    \caption{Performance comparison of ten Persian ASR models evaluated on the PSRB.}
    \label{table:asr_performance_I}
    \vspace{-7mm}
\end{table}

\section{Error Analysis}\label{sec5}

The performance of Persian ASR systems reveals a variety of error types, reflecting challenges unique to the language. Similar studies, such as Wirth \& Peinl \cite{wirth2022asr}, have categorized ASR errors across the German language. identifying issues like Minor Errors, Major Errors, Names and Loanwords, Homophones, Flawed Audio Input, and Ambiguous Audio Input. Many of these errors are also prevalent in Persian ASR. However, Persian speech recognition systems still face significant challenges due to the unique characteristics of the language, variations in speech patterns, and dataset biases. This section provides a structured analysis of the most prominent ASR errors observed in our benchmark evaluation. We categorize these errors into linguistic errors, substitution and phonetic errors, and hallucination errors presenting their impact and potential mitigation strategies.

\subsection{Linguistic Errors}\label{subsec5.1}

\subsubsection{Word Boundaries}\label{subsubsec5.1.1}

One of the most critical linguistic challenges in Persian ASR is the correct placement of word boundaries. Unlike English, Persian uses the Zero Width Non-Joiner (ZWNJ) to distinguish compound words from separate words, affecting sentence readability and grammatical correctness \cite{ghayoomi2009challenges,bijankhan2011lessons}. For example, the correct representation of trees is 
"\mypersian{درخت‌ها}"
but errors can occur when the ZWNJ is omitted or replaced, leading to misrepresentation like 
"\mypersian{درخت ها}" 
or 
"\mypersian{درختها}". 
Such missegmentations can dramatically alter the meaning of phrases, as demonstrated by 
"\mypersian{يک‌جا خريد کردم}"
(I bought everything at once) potentially being misinterpreted as "\mypersian{يک جا خريد کردم}" 
(I bought at one place). Because WER depends on proper word boundaries, these errors significantly inflate the overall error rates compared to CER, where a spacing issue only affects a single character. To mitigate these challenges, implementing post-processing normalization techniques and incorporating morphological analysis into ASR models are essential strategies.

\subsubsection{Formality}\label{subsubsec5.1.2}

As discussed in Section \ref{subsec3.1} regarding the definition of formality, Persian encompasses both formal and informal variants. ASR systems sometimes misinterpret the formality of speech, leading to inappropriate word choices or verb conjugations in transcriptions. When ASR models fail to correctly recognize and adapt to these differences, the resulting transcripts often have errors such as wrong word choices, missing words, or incorrect spellings. For example, an informal phrase like 
"\mypersian{می‌خوام برم}"
(I want to go) might be improperly transcribed into its formal equivalent 
"\mypersian{می‌خواهم بروم}" 
or vice versa. Similarly, in English ASR systems, informal contractions like "gonna" (going to) or "wanna" (want to) might be transcribed into their formal equivalents or vice versa, affecting the accuracy of the transcription. Such errors significantly increase WER and CER, as even small mistakes can change the meaning or make the text harder to understand. 

\subsubsection{He-Kasreh (\texorpdfstring{\mypersian{هِ}}{He-Kasreh})}\label{subsubsec5.1.3}

A key challenge in Persian ASR systems is the recognition of "He-Kasreh," where the letter 
"\mypersian{ه}"
is pronounced with the short vowel "/e/" to indicate possession or connection. Misrecognition of He-Kasreh can lead to syntactic errors in transcription. For example, the phrase 
"\mypersian{کتاب او}"
(his book) might be incorrectly transcribed as 
"\mypersian{کتابه او}"
introducing the grammatically incorrect He-Kasreh. Such errors distort grammatical relationships and directly impact evaluation metrics. Accurately handling He-Kasreh is essential to minimize these errors and improve the performance of Persian ASR systems.

\subsection{Interpretation Error}\label{subsec5.4}

An issue observed during the error analysis of Persian ASR systems, such as Seamless, is the generation of rephrased or paraphrased transcriptions rather than a direct match to the input speech. This error often occurs due to the model's reliance on multitask training, where it is optimized not only for transcription but also for generating semantically meaningful outputs. In such cases, the model prioritizes capturing the general intent or meaning of the input rather than preserving its exact linguistic structure.

For instance, the spoken phrase 
"\mypersian{خوش می‌گذره}"
(it's enjoyable) might be transcribed as 
"\mypersian{حال می‌ده}"
(it feels good). While the two phrases convey a similar sentiment, the substitution reflects a failure to produce an exact transcript. This issue can occur when training on datasets with paraphrased or loosely aligned text-audio pairs, leading the model to favor semantic equivalence over literal accuracy. Such errors increase WER and CER, particularly in applications where precision in transcription is critical, such as legal, medical, or academic contexts. Addressing this problem requires fine-tuning Persian ASR systems on high-quality, domain-specific datasets that emphasize exact word alignment and penalize paraphrasing tendencies during training.

\subsection{Hallucinations}\label{subsec5.5}

Hallucinations in ASR occur when models generate transcriptions that are not based on the actual audio input. These outputs may appear fluent and coherent but are factually incorrect or nonsensical \cite{frieske2024hallucinations}. The primary causes of hallucinations include poor-quality or corrupted training data, biases in ASR language models, and the inherent complexity and ambiguity of human speech. For example, an ASR system might incorrectly transcribe or generate entirely nonsensical phrases unrelated to the original speech. These errors undermine the credibility of ASR systems, potentially leading to misinterpretations and incorrect actions, particularly in high-stakes applications like medical transcription and command-based systems.

Models that incorporate semantic content modeling and multilingual ASR systems are prone to hallucination. Approaches such as Whisper, Seamless, and transducer-based decoders integrate semantic modeling into their architectures, which improves generalization but can also lead to inaccurate transcriptions that deviate from the original audio input. Furthermore, multilingual models like Whisper and Seamless may occasionally produce transcriptions in the wrong language due to errors in their language identification modules, further compromising transcription accuracy.

In this study, we perform an ablation analysis on the NVIDIA NeMo FastConformer \cite{rekesh2023fast} model for Persian(FC-Fa) to examine hallucination issues. The model features two decoding mechanisms: a CTC-based decoder \cite{graves2006connectionist} and a transducer-based decoder \cite{graves2012sequence}. The transducer decoder incorporates a predictor network that learns semantic content, effectively acting as an internal language model during training. While this architecture enhances fluency and coherence, it also increases the risk of hallucinations, particularly when the training data contains inconsistencies.

Additionally, although the overall error rates of the two decoders appear similar\footnote{https://catalog.ngc.nvidia.com/orgs/nvidia/teams/nemo/models/stt\_fa\_fastconformer\_hybrid\_large}, our proposed benchmark reveals a significant performance gap between them. Analyzing the outputs of both decoders, we observe instances where their transcriptions are identical, while in other cases, there are substantial discrepancies. As shown in the Table \ref{table:decoder_ablation}, the CTC decoder exhibits considerably greater robustness than the transducer decoder. Therefore, in this study, the results obtained from the CTC decoder for this model are reported.

\begin{table}[htbp]
    \centering
    \renewcommand{\arraystretch}{1.2}
    \begin{tabular}{lcccc}
        \toprule
        \textbf{Model} & \textbf{Decoder} & \textbf{CER} & \textbf{WER} & \textbf{SW-WER} \\
        \midrule
        \multirow{2}{*}{FC-Fa} & CTC        & 19.42 & 44.85 & 37.58 \\
                               & Transducer & 40.73 & 58.03 & 50.18 \\
        \bottomrule
    \end{tabular}
    \caption{Ablation study on decoder type of FastConformer model}
    \label{table:decoder_ablation}
    \vspace{-9mm}
\end{table}

To reduce hallucinations in ASR systems, it is essential to use high-quality training data, advanced evaluation metrics, noise-resistant models, and effective post-processing techniques. Training on diverse datasets with accurate annotations helps models establish correct associations between speech and text. Beyond Word Error Rate, metrics such as Hallucination Error Rate offer a more refined evaluation of transcription accuracy \cite{atwany2025lost}. Improving noise robustness by exposing models to varied acoustic environments enhances their stability \cite{atwany2025lost}. Post-processing methods, including Voice Activity Detection (VAD), help filter out non-speech segments that may contribute to hallucinations \cite{baranski2025investigation}. Additionally, semantic consistency checks ensure that transcriptions align with the original speech content \cite{frieske2024hallucinations}. These efforts aim to improve ASR reliability, ensuring that transcriptions remain accurate and trustworthy in critical applications.

\section{Robustness}\label{sec6}

In this section, we analyze the robustness of ASR systems, with results presented in Table \ref{table:Pvote_formal_noisy}, the CER, WER, and SW-WER breakdown for Formality and noisiness. Based on the results, the error rate is higher in noisy conditions. Also, when the speaker speaks formally, the error rate is lower than in informal conditions. As shown in Table \ref{table:Pvote_formal_noisy}, the Avanegar model generally performs well across different conditions; however, in the informal mode, the Chirp model achieves the best performance among all models, particularly in the informal noisy setting. In contrast, the Avanegar model outperforms others in the formal clean, formal noisy, and informal clean conditions. Interestingly, the Faster Whisper model surpasses Azure in performance within the informal noisy condition, highlighting its relative robustness in handling challenging speech scenarios.

Additionally, open-source models tend to have lower performance compared to commercial models, with the Faster Whisper model demonstrating the best results among open-source alternatives. Notably, ASR models appear to be more robust to noise than to informality in the input speech. This is because noise robustness can be effectively addressed through data augmentation techniques, whereas handling informality involves differences in both speech and text structure. As demonstrated in Section 5, the linguistic differences between informal and formal Persian are substantial. Achieving robustness to variations in formality necessitates real spoken data that accurately represents diverse speech styles, along with precise transcription.

\begin{table*}[htbp]
    \centering
    \small
    \renewcommand{\arraystretch}{1.2}
    \resizebox{\textwidth}{!}{
    \begin{tabular}{l lccc ccc ccc}
        \toprule
        \multirow{2}{*}{\textbf{ }} & \multirow{2}{*}{\textbf{Models}} & \multicolumn{3}{c}{Formal} & \multicolumn{3}{c}{Informal} & \multicolumn{3}{c}{Avg} \\
        \cmidrule(lr){3-5} \cmidrule(lr){6-8} \cmidrule(lr){9-11}
        & & \textbf{CER} & \textbf{WER} & \textbf{SW-WER} & \textbf{CER} & \textbf{WER} & \textbf{SW-WER} & \textbf{CER} & \textbf{WER} & \textbf{SW-WER} \\
        \midrule
        \multirow{10}{*}{\rotatebox[origin=c]{90}{\textbf{Clean}}} 
        & Vosk & 12.2 & 26.8 & 20.5 & 22.3 & 43.7 & 32.7 & 19.1 & 38.3 & 28.8 \\
        & Seamless & 10.4 & 22.5 & 16.9 & 21.8 & 38.7 & 29.7 & 18.2 & 33.6 & 25.6 \\
        & Whisper & 12.7 & 30.0 & 20.3 & 17.0 & 37.9 & 27.4 & 15.6 & 35.3 & 25.2 \\
        & Faster-Whisper & 8.1 & 26.3 & 16.3 & 13.1 & 32.4 & 21.8 & 11.5 & 30.5 & 20.1 \\
        & FC-Fa & 11.3 & 33.6 & 22.9 & 18.8 & 43.8 & 31.0 & 16.4 & 40.5 & 28.4 \\
        & SLPL W2V2 & 10.5 & 35.4 & 23.6 & 17.8 & 44.1 & 30.3 & 15.5 & 41.3 & 28.2 \\
        & Azure & 7.1 & 18.4 & 13.6 & 16.6 & 36.3 & 25.8 & 13.6 & 30.6 & 21.9 \\
        & Chirp & 5.2 & 14.0 & 10.1 & 8.9 & 19.2 & 14.1 & 7.8 & 17.5 & 12.8 \\
        & Aipa & 6.5 & 18.5 & 13.2 & 10.0 & 23.7 & 17.2 & 8.9 & 22.1 & 15.9 \\
        & Avanegar & \textbf{4.4} & \textbf{11.5} & \textbf{8.7} & \textbf{8.3} & \textbf{18.6} & \textbf{13.4} & \textbf{7.0} & \textbf{16.4} & \textbf{11.9} \\
        \midrule
        \multirow{10}{*}{\rotatebox[origin=c]{90}{\textbf{Noisy}}} 
        & Vosk & 15.2 & 30.6 & 23.0 & 32.9 & 57.2 & 44.8 & 28.4 & 50.3 & 39.1 \\
        & Seamless & 13.6 & 27.1 & 21.0 & 30.3 & 49.3 & 39.3 & 26.0 & 43.6 & 34.6 \\
        & Whisper & 14.6 & 37.6 & 27.9 & 24.4 & 50.3 & 37.9 & 21.9 & 47.0 & 35.3 \\
        & Faster-Whisper & 10.2 & 29.0 & 19.2 & 17.6 & 39.8 & 27.2 & 15.7 & 37.0 & 25.1 \\
        & FC-Fa & 14.4 & 37.4 & 25.8 & 24.8 & 52.6 & 37.8 & 22.1 & 48.7 & 34.7 \\
        & SLPL W2V2 & 13.9 & 40.5 & 27.4 & 25.2 & 55.3 & 39.2 & 22.1 & 51.6 & 36.8 \\
        & Azure & 7.9 & 19.8 & 14.4 & 21.4 & 39.7 & 31.1 & 17.8 & 37.0 & 26.8 \\
        & Chirp & 6.9 & 16.2 & 12.0 & \textbf{11.4} & \textbf{24.1} & \textbf{17.4} & \textbf{10.2} & 22.1 & 16.0 \\
        & Aipa & 7.5 & 20.5 & 14.5 & 13.3 & 29.2 & 21.1 & 11.8 & 27.0 & 19.4 \\
        & Avanegar & \textbf{5.1} & \textbf{12.6} & \textbf{9.3} & 12.1 & 25.1 & 18.2 & 10.3 & \textbf{21.9} & \textbf{15.9} \\
        \midrule
        \multirow{10}{*}{\rotatebox[origin=c]{90}{\textbf{Average}}} 
        & Vosk & 13.6 & 28.4 & 21.7 & 28.1 & 51.1 & 39.3 & 24.0 & 44.8 & 32.8 \\
        & Seamless & 11.9 & 24.7 & 18.8 & 26.8 & 44.5 & 35.0 & 22.1 & 38.5 & 30.6 \\
        & Whisper & 13.6 & 33.3 & 23.9 & 21.1 & 44.7 & 33.2 & 18.5 & 40.5 & 30.5 \\
        & Faster-Whisper & 9.1 & 27.6 & 19.1 & 15.3 & 36.5 & 25.6 & 13.4 & 33.4 & 23.2 \\
        & FC-Fa & 12.8 & 35.0 & 24.1 & 21.8 & 48.3 & 34.8 & 19.1 & 43.2 & 31.0 \\
        & SLPL W2V2 & 12.1 & 37.8 & 25.4 & 21.9 & 51.1 & 36.9 & 19.1 & 46.7 & 32.3 \\
        & Azure & 7.5 & 19.1 & 13.9 & 19.1 & 38.5 & 28.7 & 15.1 & 32.8 & 24.3 \\
        & Chirp & 6.0 & 15.1 & 11.0 & \textbf{10.3} & \textbf{21.9} & \textbf{15.9} & 9.0 & 19.9 & 14.5 \\
        & Aipa & 7.0 & 19.5 & 13.8 & 12.8 & 26.7 & 19.4 & 10.4 & 24.6 & 17.3 \\
        & Avanegar & \textbf{4.7} & \textbf{12.0} & \textbf{9.0} & 10.4 & 22.2 & 16.0 & \textbf{8.7} & \textbf{19.3} & \textbf{14.0} \\
        \bottomrule
    \end{tabular}
    }
    \caption{Evaluation of ten ASR models across formal and informal speech in clean and noisy conditions. Metrics are reported as percentages (\%).}
    \label{table:Pvote_formal_noisy}
    \vspace{-5mm}
\end{table*}

\subsection{Number of Speakers Effect}\label{subsec6.1}

Table \ref{table:speaker_robustness} shows that the error rate in processing multi-speaker audio is higher than in single-speaker mode across all models, both commercial and open-source. This variation depends on factors such as model architecture, training data, and training methods. The primary challenge in multi-speaker scenarios is overlapping speech, which complicates transcription.  In the benchmark, efforts were made to minimize speech overlap as much as possible, ensuring that speakers did not talk simultaneously. While SSL models perform well in single-speaker scenarios, they struggle with multi-speaker inputs, often resulting in high deletion errors and sometimes omitting an entire speaker from the transcript. SSL models are designed to generate highly informative representations of input speech, creating embeddings that are crucial for various speech processing downstream tasks, such as speech recognition, speaker verification, and speech translation. To optimize performance in speaker-related tasks, these models are typically trained on single-speaker utterances, allowing them to learn speaker representations and enhance robustness. However, to effectively perform speech recognition in multi-talker scenarios, they must be adapted accordingly. 

\begin{table*}[htbp]
    \centering
    \renewcommand{\arraystretch}{1.2}
    \resizebox{0.9\textwidth}{!}{
    \begin{tabular}{lccc|ccc}
        \hline
        \multirow{2}{*}{\textbf{Model}} & \multicolumn{3}{c|}{\textbf{Single Speaker}} & \multicolumn{3}{c}{\textbf{Multiple Speaker}} \\
        & \textbf{CER} & \textbf{WER} & \textbf{SW-WER} & \textbf{CER} & \textbf{WER} & \textbf{SW-WER} \\
        \hline
        Vosk & 21.0 & 40.9 & 30.9 & 40.7 & 66.4 & 53.8 \\
        Seamless & 18.2 & 34.1 & 25.9 & 45.8 & 66.2 & 56.1 \\
        Whisper & 16.6 & 38.4 & 28.0 & 32.3 & 59.1 & 45.6 \\
        Faster-Whisper & 11.7 & 31.3 & 23.0 & 23.9 & 51.9 & 35.5 \\
        SLPL W2V2 & 16.3 & 43.4 & 29.3 & 34.9 & 66.1 & 49.3 \\
        FC-Fa & 16.8 & 41.8 & 28.9 & 34.5 & 62.5 & 48.0 \\
        Azure & 13.4 & 31.1 & 21.9 & 29.4 & 50.3 & 39.4 \\
        Chirp & 7.3 & 17.6 & 12.6 & 19.1 & \textbf{33.2} & \textbf{25.6} \\
        Aipa & 8.6 & 22.5 & 15.9 & 20.6 & 36.8 & 28.2 \\
        Avanegar & \textbf{7.0} & \textbf{16.8} & \textbf{11.9} & \textbf{18.9} & 34.1 & 25.7 \\
        \hline
    \end{tabular}
    }
    \caption{Comparison of ASR model performance (in \%) in single-speaker versus multi-speaker scenarios}
    \label{table:speaker_robustness}
    \vspace{-5mm}
\end{table*}

For example, in the single speaker mode, the SLPL model, which is based on the wav2vec2 architecture, performs identically to the whisper model, while in the multi-speaker mode, the whisper model performs better in all evaluation metrics. As another example, the seamless model has lower WER and SW-WER than the whisper model in the single-speaker case, while in the multi-speaker case, the whisper model outperforms by a significant margin. 

To address this issue, approaches such as speech separation, end-to-end multi-talker automatic speech recognition (ASR) models, and target-speaker ASR have been explored in recent years  \cite{huang2023adapting,meng2024empowering,li2023adapting}. However, multi-speaker ASR remains a challenging task, particularly for Persian ASR systems, where the error rate in multi-speaker settings is consistently higher than in single-speaker mode.

\subsection{Bias Against Age}\label{subsec6.2}

\begin{figure}[htbp]
    \centering
    
    \begin{subfigure}[b]{0.32\textwidth}
        \centering
        \includegraphics[width=\linewidth]{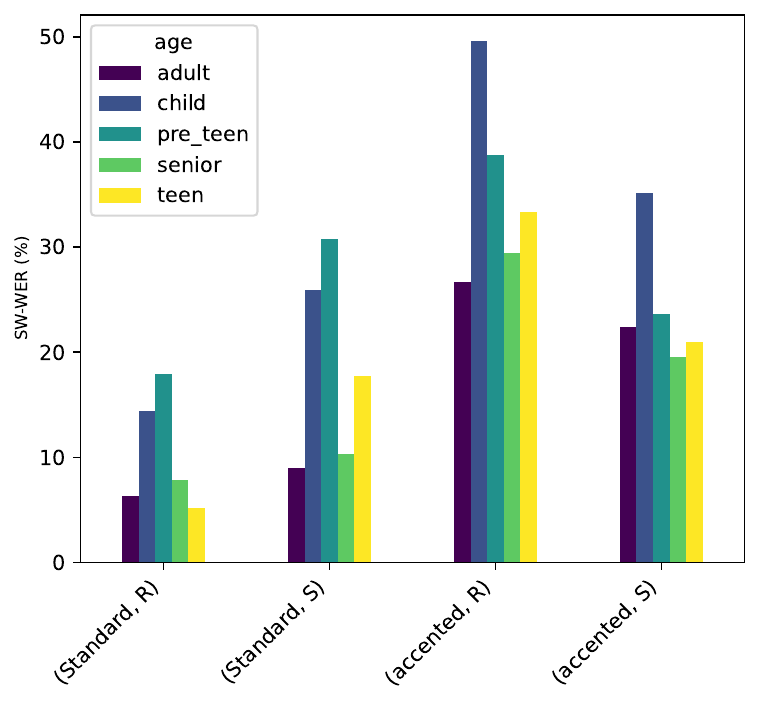}
        \caption{Avanegar}
    \end{subfigure}
    \begin{subfigure}[b]{0.32\textwidth}
        \centering
        \includegraphics[width=\linewidth]{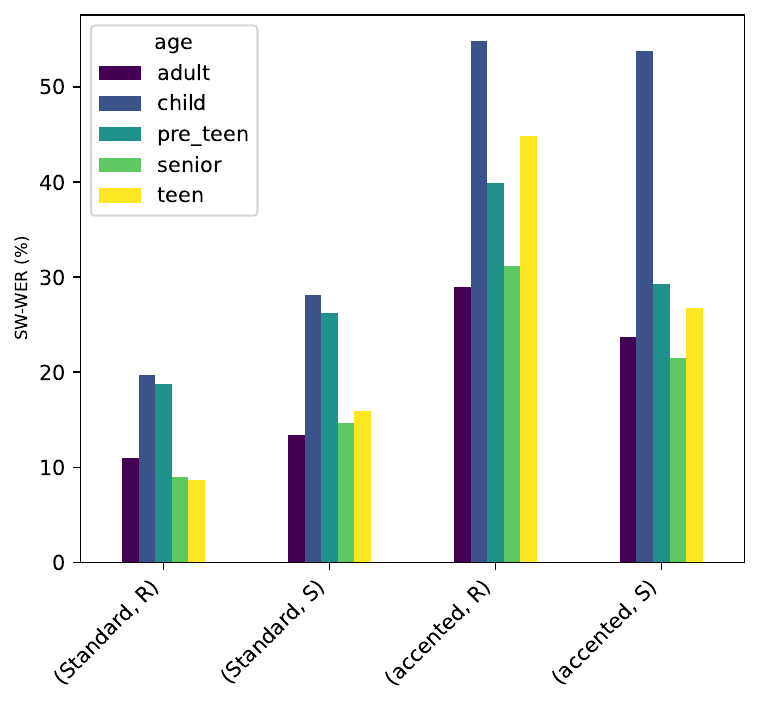}
        \caption{Aipa}
    \end{subfigure}
    \begin{subfigure}[b]{0.32\textwidth}
        \centering
        \includegraphics[width=\linewidth]{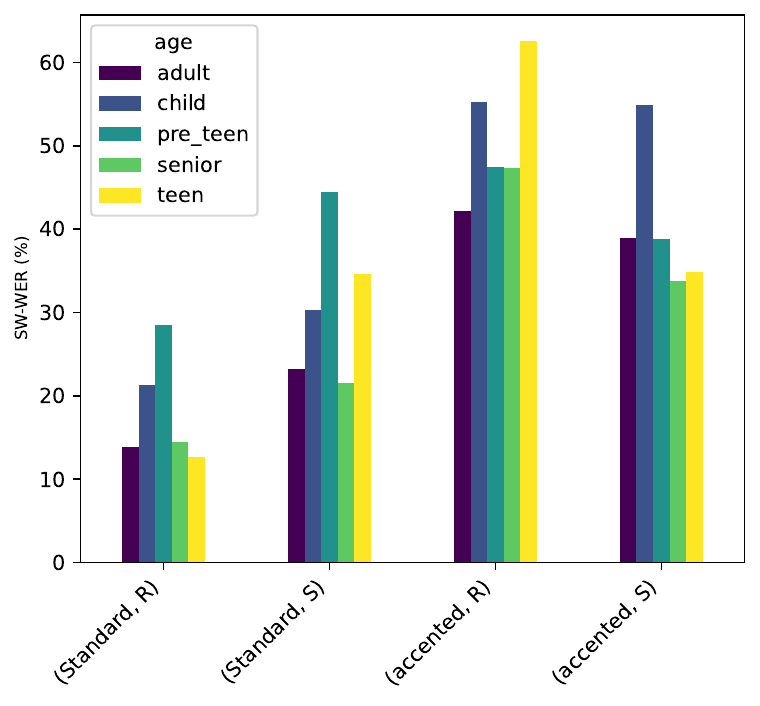}
        \caption{Seamless}
    \end{subfigure}
    
    \vspace{0.2cm} 
    
    \begin{subfigure}[b]{0.32\textwidth}
        \centering
        \includegraphics[width=\linewidth]{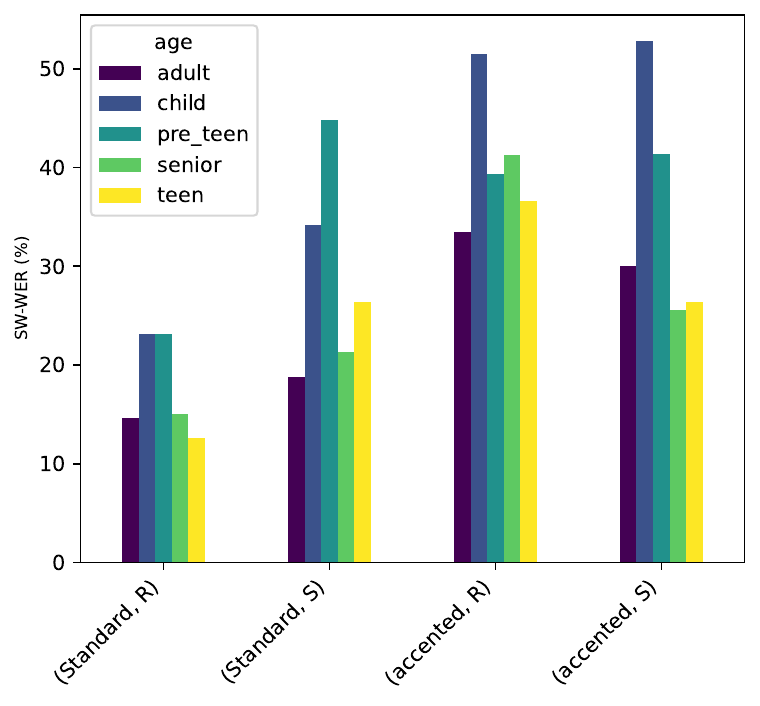}
        \caption{Faster-Whisper}
    \end{subfigure}
    \begin{subfigure}[b]{0.32\textwidth}
        \centering
        \includegraphics[width=\linewidth]{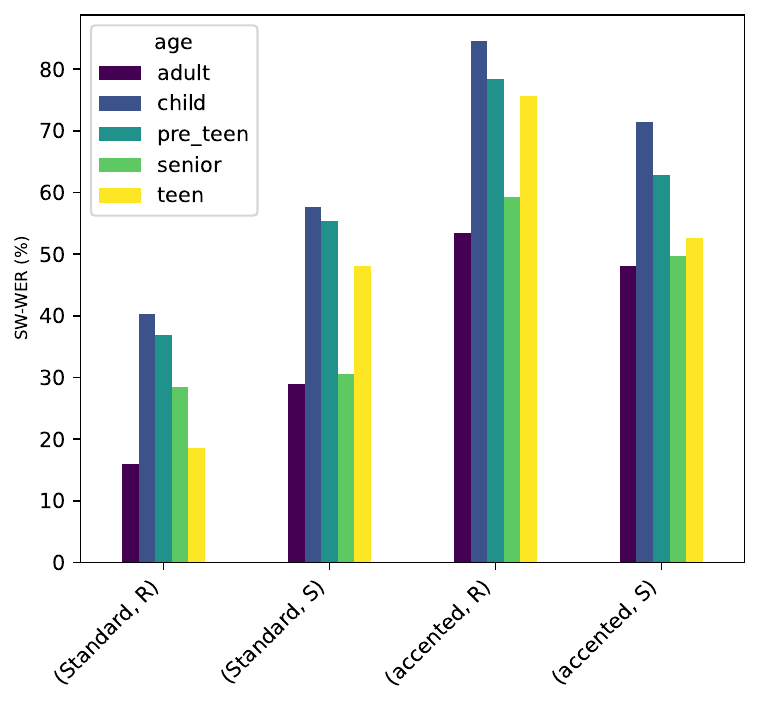}
        \caption{Vosk}
    \end{subfigure}
    \begin{subfigure}[b]{0.32\textwidth}
        \centering
        \includegraphics[width=\linewidth]{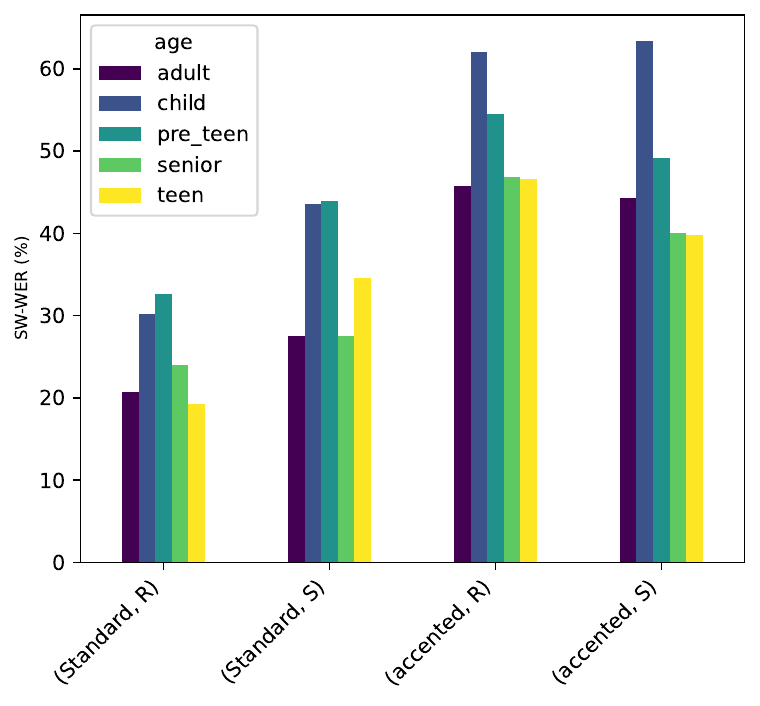}
        \caption{FC-Fa}
    \end{subfigure}
    
    \vspace{0.2cm}
    
    \begin{subfigure}{0.32\textwidth}
        \centering
        \includegraphics[width=\linewidth]{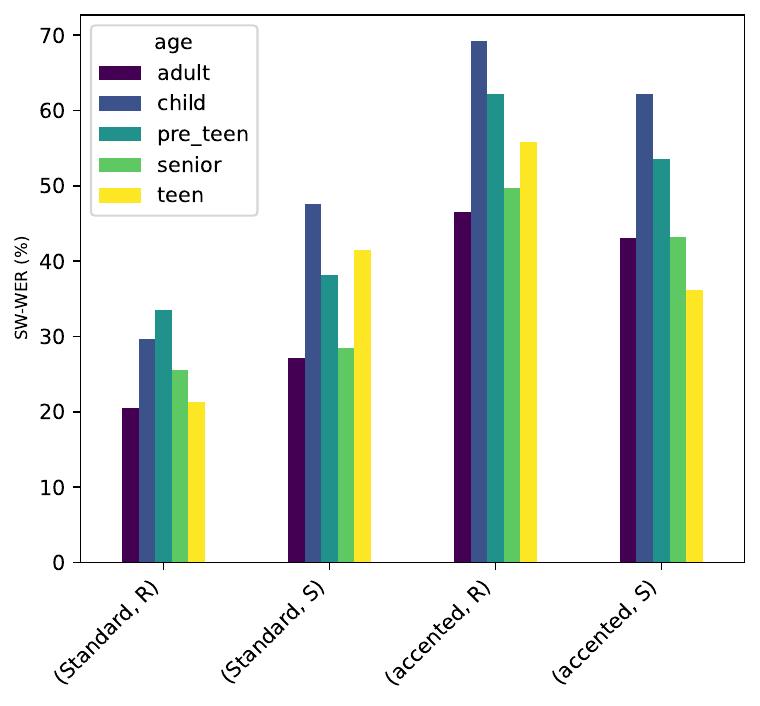}
        \caption{SLPL W2V2}
    \end{subfigure}
    \begin{subfigure}{0.32\textwidth}
        \centering
        \includegraphics[width=\linewidth]{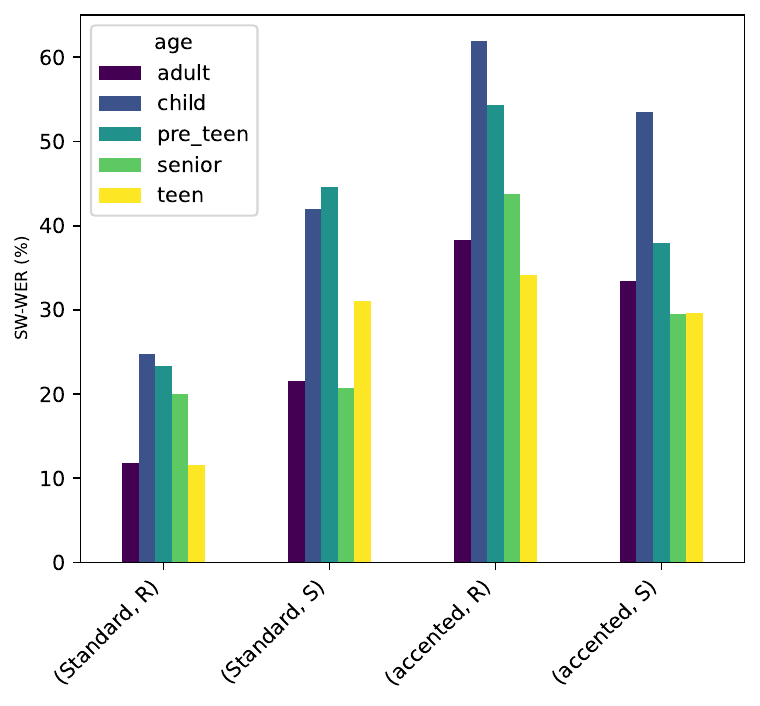}
        \caption{Azure}
    \end{subfigure}
    \begin{subfigure}{0.32\textwidth}
        \centering
        \includegraphics[width=\linewidth]{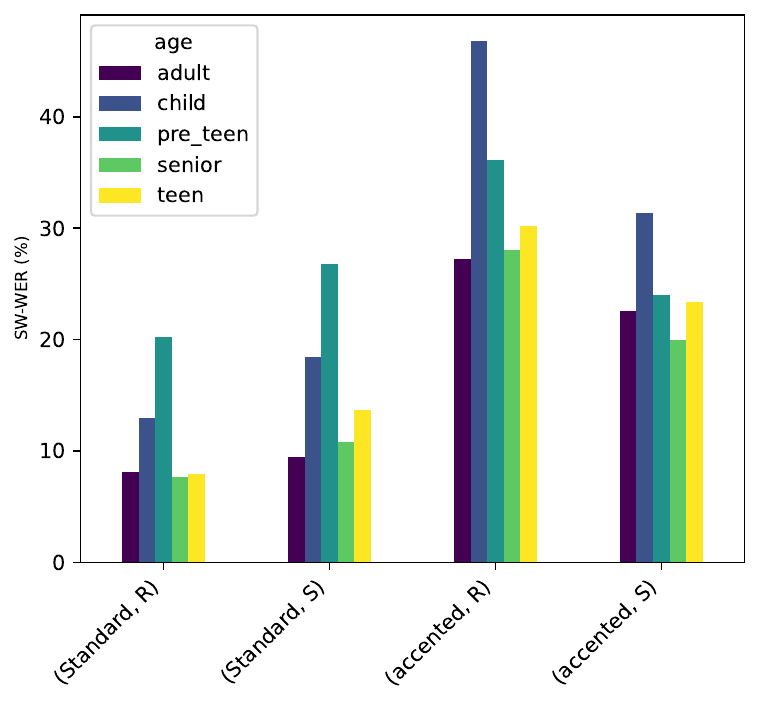}
        \caption{Chirp}
    \end{subfigure}
    
    \caption{Bar plots showing SW-WER across age groups (children, teens, adults, seniors), spontaneity levels (read(R) vs. spontaneous speech(S)), and accent categories (standard vs. accented) for nine ASR models}
    \label{fig:ageplot}
    \vspace{-6mm}
\end{figure}

Overall, in both Standard Persian and accented Persian, adult speakers have the best output for all models. This is followed by senior and teen speakers. Child speakers are recognized the worst (see Figure \ref{fig:ageplot} for the SW-WER breakdown for age, spontaneity, and accent).  

The disparity in performance between ASR models for adult and children's speech is primarily due to the fact that most models are trained on datasets predominantly composed of adult speech, resulting in mismatches when processing children's voices. Crowdsourcing audio data collection can exacerbate this issue, as it often introduces bias toward adult and senior speakers, leading to a lack of robustness in ASR models across different age groups, particularly for children. To create a model that performs reliably across all age groups, it is essential to collect data representing a diverse range of ages. However, this approach can be costly due to the significant expenses associated with audio data labeling and related processes.

To address these challenges, recent studies have explored techniques such as speaker normalization and adaptation algorithms aimed at reducing acoustic mismatches between children's and adults' speech. Despite these advancements, ASR models still exhibit lower robustness when processing children's speech \cite{bhardwaj2022automatic}. This is partly because children's speech presents unique challenges, including anatomical and developmental differences in the vocal tract, which result in distinct acoustic and linguistic properties. Factors such as higher pitch, greater variability in pronunciation, and less stable prosodic features make children's speech more difficult to model effectively. Furthermore, the limited availability of large-scale, high-quality children's speech datasets further hampers ASR performance.

\subsection{Gender Effect}\label{subsec6.3}

\begin{figure}[htbp]
    \centering
    
    \begin{subfigure}[b]{0.32\textwidth}
        \centering
        \includegraphics[width=\linewidth]{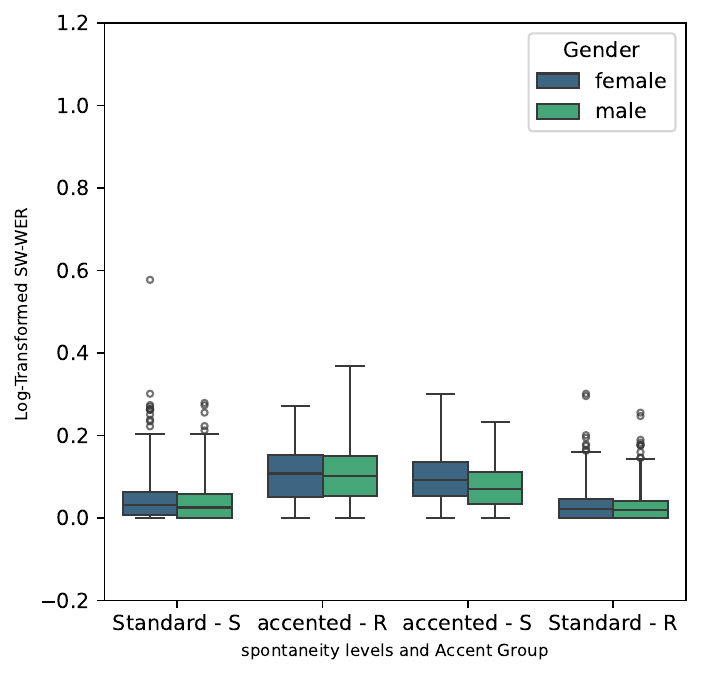}
        \caption{Avanegar}
    \end{subfigure}
    \begin{subfigure}[b]{0.32\textwidth}
        \centering
        \includegraphics[width=\linewidth]{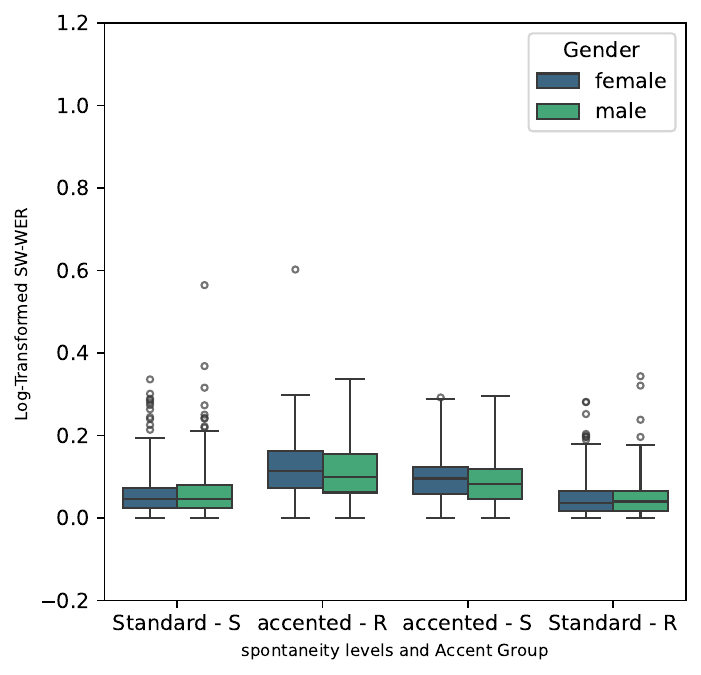}
        \caption{Aipa}
    \end{subfigure}
    \begin{subfigure}[b]{0.32\textwidth}
        \centering
        \includegraphics[width=\linewidth]{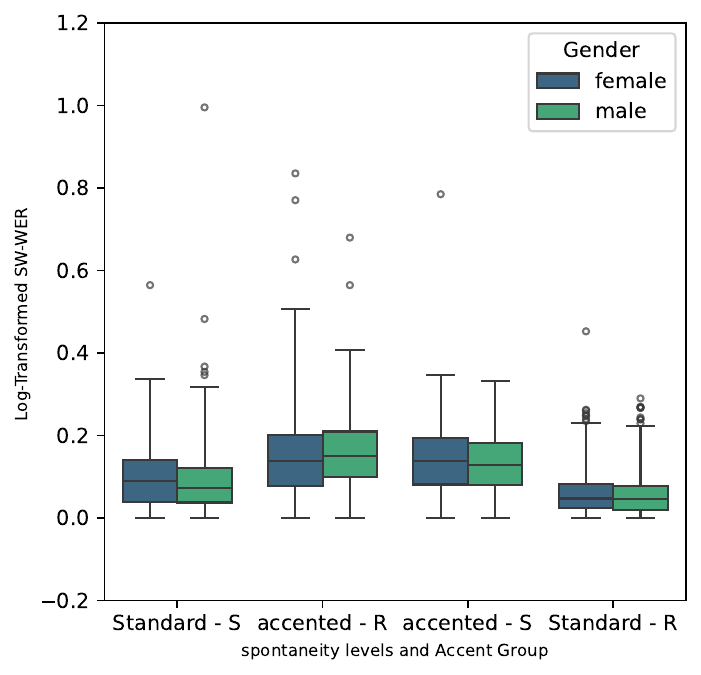}
        \caption{Seamless}
    \end{subfigure}
    
    \vspace{0.2cm} 
    
    \begin{subfigure}[b]{0.32\textwidth}
        \centering
        \includegraphics[width=\linewidth]{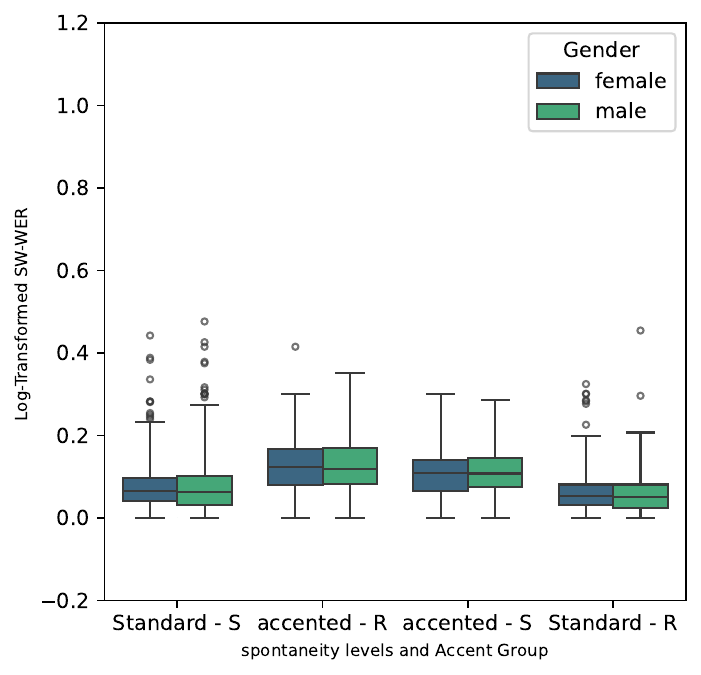}
        \caption{Faster-Whisper}
    \end{subfigure}
    \begin{subfigure}[b]{0.32\textwidth}
        \centering
        \includegraphics[width=\linewidth]{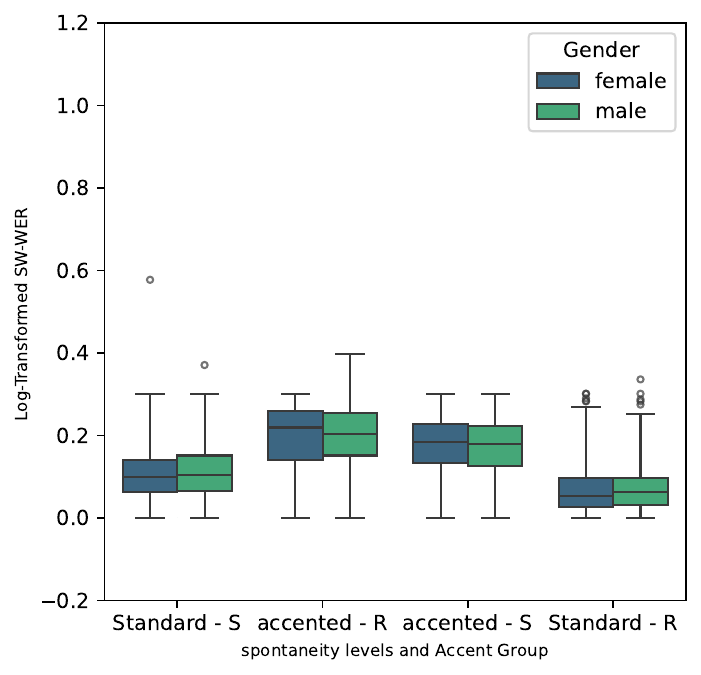}
        \caption{Vosk}
    \end{subfigure}
    \begin{subfigure}[b]{0.32\textwidth}
        \centering
        \includegraphics[width=\linewidth]{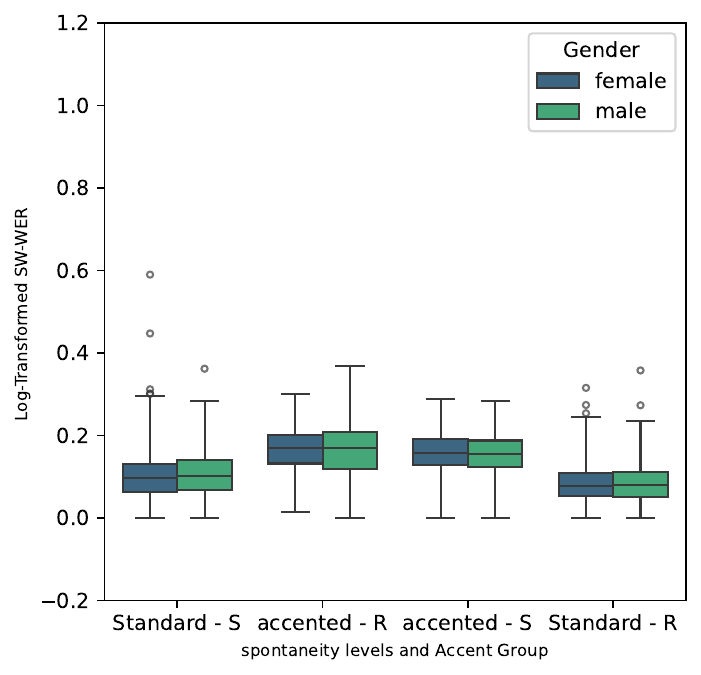}
        \caption{FC-Fa}
    \end{subfigure}
    
    \vspace{0.2cm}
    
    \begin{subfigure}{0.32\textwidth}
        \centering
        \includegraphics[width=\linewidth]{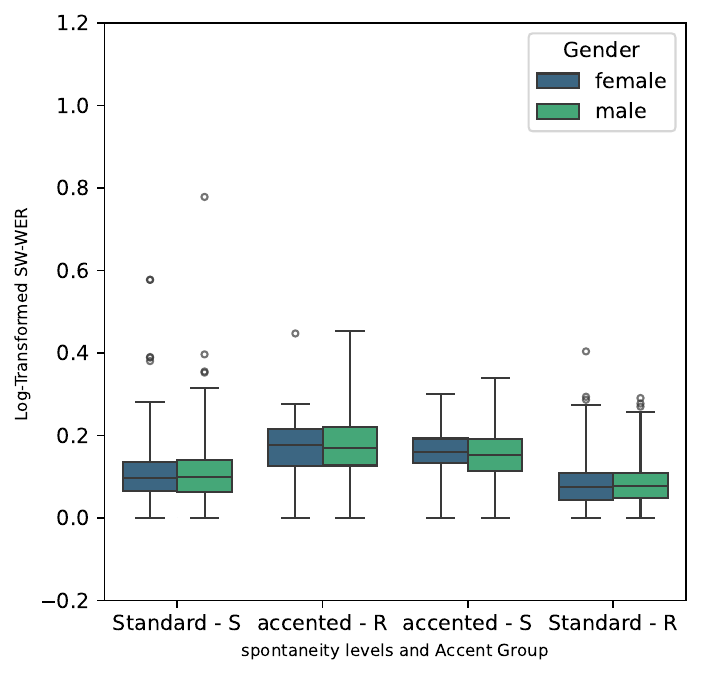}
        \caption{SLPL W2V2}
    \end{subfigure}
    \begin{subfigure}{0.32\textwidth}
        \centering
        \includegraphics[width=\linewidth]{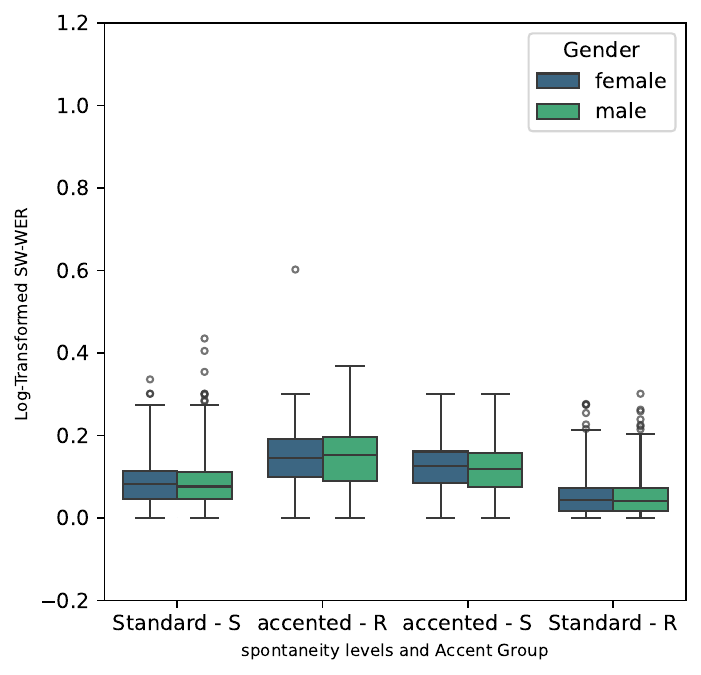}
        \caption{Azure}
    \end{subfigure}
    \begin{subfigure}{0.32\textwidth}
        \centering
        \includegraphics[width=\linewidth]{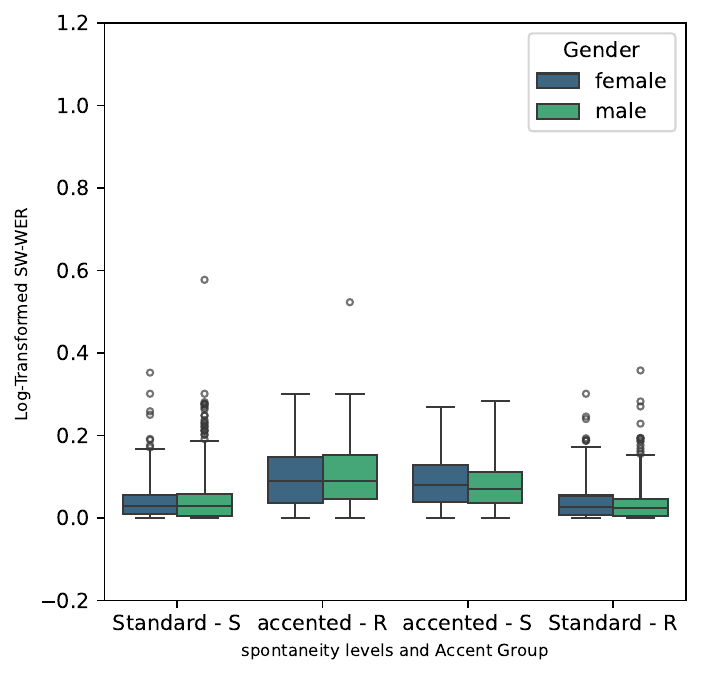}
        \caption{Chirp}
    \end{subfigure}
    
    \caption{Box plots on a logarithmic scale depicting SW-WER variations across gender, spontaneity, and accent for nine ASR models.}
    \label{fig:genderbias}
    \vspace{-5mm}
\end{figure}

For the gender analysis, spontaneity and accent were considered, and based on the obtained results, box plots were generated for these factors (see Figure \ref{fig:genderbias}). The chosen evaluation metric is the SW-WER, as it is a robust measure for assessing the model's actual performance in accent variations. To enhance visualization, a logarithmic scale was applied to the vertical axis, and outlier data points were excluded from the analysis.

Our analysis reveals a very slight bias in ASR models with respect to gender. As observed in the Figure \ref{fig:genderbias}, recognition accuracy is slightly lower for male speakers in accented speech. This aligns with findings from previous research \cite{fuckner2023uncovering}, which demonstrated that female speech is generally recognized more accurately than male speech across most models. This discrepancy may be due to differences in acoustic features between accented and standard speech, which are more pronounced in male speakers due to their lower pitch.

\subsection{Bias Against Regional Accents}\label{subsec6.4}

Persian is spoken across various regions, each with a different accent that differs in vocabulary, grammar, and sentence structure. However, most ASR models are trained and evaluated primarily on Standard Persian, leading to a bias against regional accents.

The more complex the specialized vocabulary and linguistic structure of a given accent, the greater the challenges faced by speech-to-text models, leading to higher error rates. Given the increased likelihood of substitution errors during model evaluation, the SW-WER metric was employed to provide a more precise and robust assessment of each model's performance.

In this study, 12 accents were evaluated: Baluchi, Dari, Isfahani, Jonubi, Kermani, Kurdish, Lori, Mashhadi, Shirazi, Shomali, Turkish, and Yazdi, along with Standard Persian. The results indicate that ASR systems consistently perform worse on accented speech compared to Standard Persian. Among the evaluated accents, recognition accuracy was lowest for Mashhadi and Yazdi, while Turkish and Dari showed relatively better performance. These observations can be explained by an imbalance in training data—stemming from variations in the number of speakers per dialect—which limits the representation of certain regional accents during model training. Moreover, the inherent phonetic and acoustic variability, coupled with distinct linguistic and phonological features, further compounds recognition challenges and contributes to the lower accuracy observed in some dialects. Table \ref{table:accents} presents the SW-WER for each accent.

\begin{table*}[htbp]
    \centering
    \normalsize
    \renewcommand{\arraystretch}{1.3} 
    \resizebox{\textwidth}{!}{
    \begin{tabular}{l|cccccccccc|c}
        \multicolumn{1}{c|}{\diagbox{Accent}{Model}} & \textbf{Vosk} & \textbf{Seamless} & \textbf{Whisper} & \textbf{F-Whisper} & \textbf{SLPL} & \textbf{FC-Fa} & \textbf{Azure} & \textbf{Chirp} & \textbf{Aipa} & \textbf{Avanegar} & \textbf{Mean}\\
        \hline
        Baluchi & 54.7 & 51.6 & 78.7 & 32.4 & 51 & 49.9 & 40.6 & 31.6 & 27.2 & \textbf{25} & 44.2 \\
        Dari & 37.8 & 29.7 & 20.8 & 20.6 & 34.5 & 34.7 & 23.2 & 15.8 & 21.2 & \textbf{15.7} & 25.4\\
        Isfahani & 52.2 & 53.5 & 45.2 & 31.5 & 43.3 & 43.6 & 33.3 & 23 & 27.8 & \textbf{22.6} & 37.6\\
        Jonubi & 57.1 & 49.5 & 46.7 & 35.9 & 48.3 & 51.3 & 38.7 & \textbf{25.8} & 28.7 & 28.1 & 41\\
        Kermani & 54.7 & 33.3 & 30.6 & 27.8 & 41.2 & 40.9 & 33.1 & \textbf{17.8} & 22 & 19.2 & 32\\
        Kurdish & 44.6 & 34.6 & 66.6 & 32.5 & 42.7 & 39.7 & 35.6 & 23.9 & 27.9 & \textbf{19} & 36.7\\
        Lori & 50.9 & 41.6 & 56.5 & 32 & 44.4 & 43.4 & 38.7 & 27 & 32.6 & \textbf{26.2} & 39.3\\
        Mashhadi & 70 & 68.5 & 53.9 & 46.5 & 61.2 & 58.3 & 52.6 & \textbf{40.1} & 42.5 & 40.2 & 53.3\\
        Shirazi & 60.5 & 48.1 & 65.1 & 38.2 & 51.6 & 49.3 & 43.5 & \textbf{24.1} & 29.4 & 29 & 43.8\\
        Shomali & 57.7 & 48.8 & 46.6 & 35.7 & 50.6 & 48.4 & 42.6 & \textbf{29.7} & 28.2 & 30.4 & 41.8\\
        Standard & 27 & 23.9 & 24 & 18.7 & 26.9 & 26.5 & 19.5 & 10.2 & 13.7 & \textbf{9.6} & 20\\
        Turkish & 45.7 & 29.8 & 23.6 & 23.1 & 37 & 37.2 & 25 & \textbf{13.7} & 17.8 & 16.4 & 26.9\\
        Yazdi & 62.2 & 66.6 & 56.3 & 41.2 & 53.3 & 52.3 & 45.1 & 35.8 & 36.6 & \textbf{31.8} & 48.1\\
    \end{tabular}
}
    \caption{ASR Models Performance Across Different Accents (All metrics are in percentage).}
    \label{table:accents}
    \vspace{-3mm}
\end{table*}

\subsection{Data Source and Semantic Content Effect}\label{subsec6.5}

Analyzing the SW-WER of different ASR models across data sources like film, documentary, audiobook, talk show, etc, can reveal how well each model handles varied linguistic characteristics and acoustic conditions. Comparing these error rates helps identify which models generalize better to each data source and indicates the nature and biases of their training data. By identifying these disparities, we can implement targeted data augmentation to improve model robustness across diverse scenarios.

Our results show significant variability in SW-WER across the data sources. Audiobooks, characterized by clear enunciation and controlled recording conditions, consistently yielded the lowest error rates, indicating that the models are well-tuned for speech with minimal background noise and structured delivery. In contrast, talk shows, Animations, and films where spontaneous dialogue, overlapping speech, and background noise are prevalent exhibited substantially higher SW-WER values. Documentaries and News, which often include domain-specific terminology and a mix of controlled narration and ambient sounds, presented intermediate performance between these two extremes. 

\begin{figure}[!htb]
    \centering
    
    \begin{subfigure}[b]{0.48\textwidth}
        \centering
        \includegraphics[width=\linewidth]{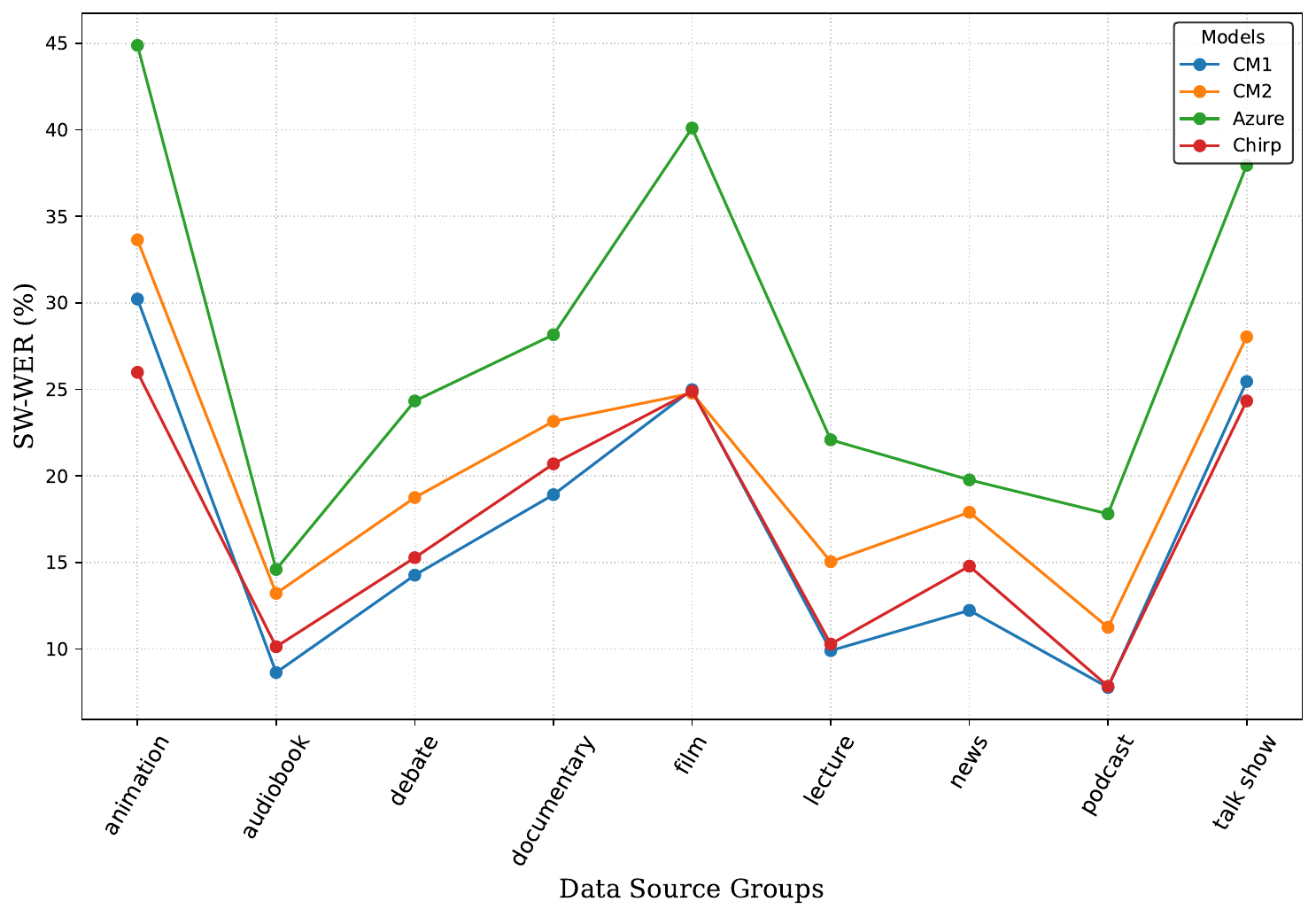}
        \caption{Commercial models}
    \end{subfigure}
    \begin{subfigure}[b]{0.48\textwidth}
        \centering
        \includegraphics[width=\linewidth]{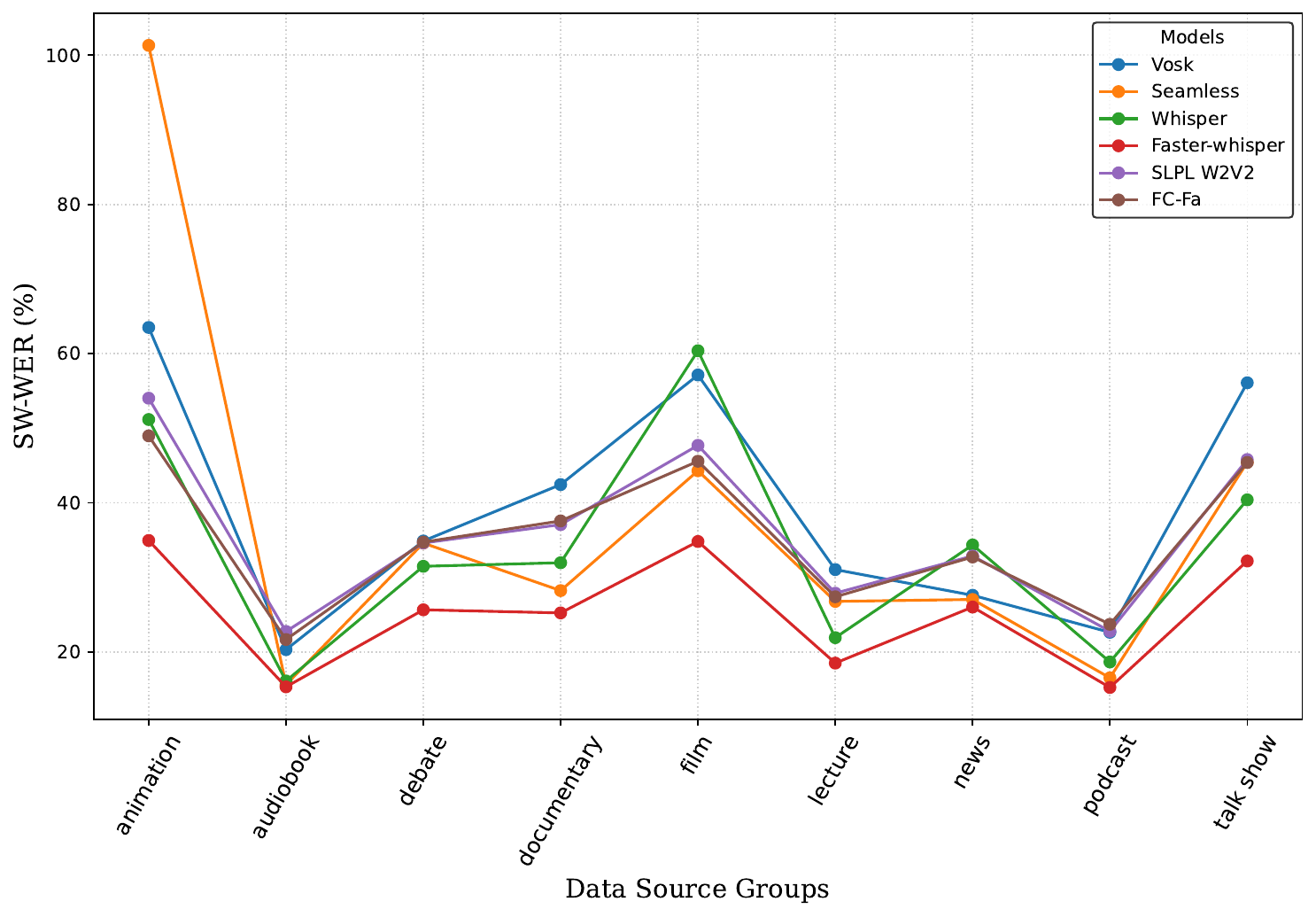}
        \caption{Open-source models}
    \end{subfigure}
    \caption{ASR models’ SW-WER across data sources (e.g., audiobooks, films, talk shows), showing lower errors for audiobooks and higher for sources like talk shows.}
    \label{fig:datasources}
    \vspace{-5mm}
\end{figure}

Figure \ref{fig:datasources} indicates that the Avanegar and Aipa  models demonstrate greater robustness compared to other models when handling various data sources. In contrast, open-source and multilingual models like Whisper and Seamless struggle with some areas, including film, talk shows, and animation.

Models may perform variably based on their exposure to domain-specific terminology and language structures during training. For example, a model that has been mainly trained on general conversational data might struggle with the specialized vocabulary and complex sentence structures found in medical or financial texts, leading to higher error rates. Conversely, a model fine-tuned on a particular domain may show lower SW-WER in that area but might underperform in others where it lacks sufficient training examples.

\begin{figure}[!htb]
    \centering
    
    \begin{subfigure}[b]{0.48\textwidth}
        \centering
        \includegraphics[width=\linewidth]{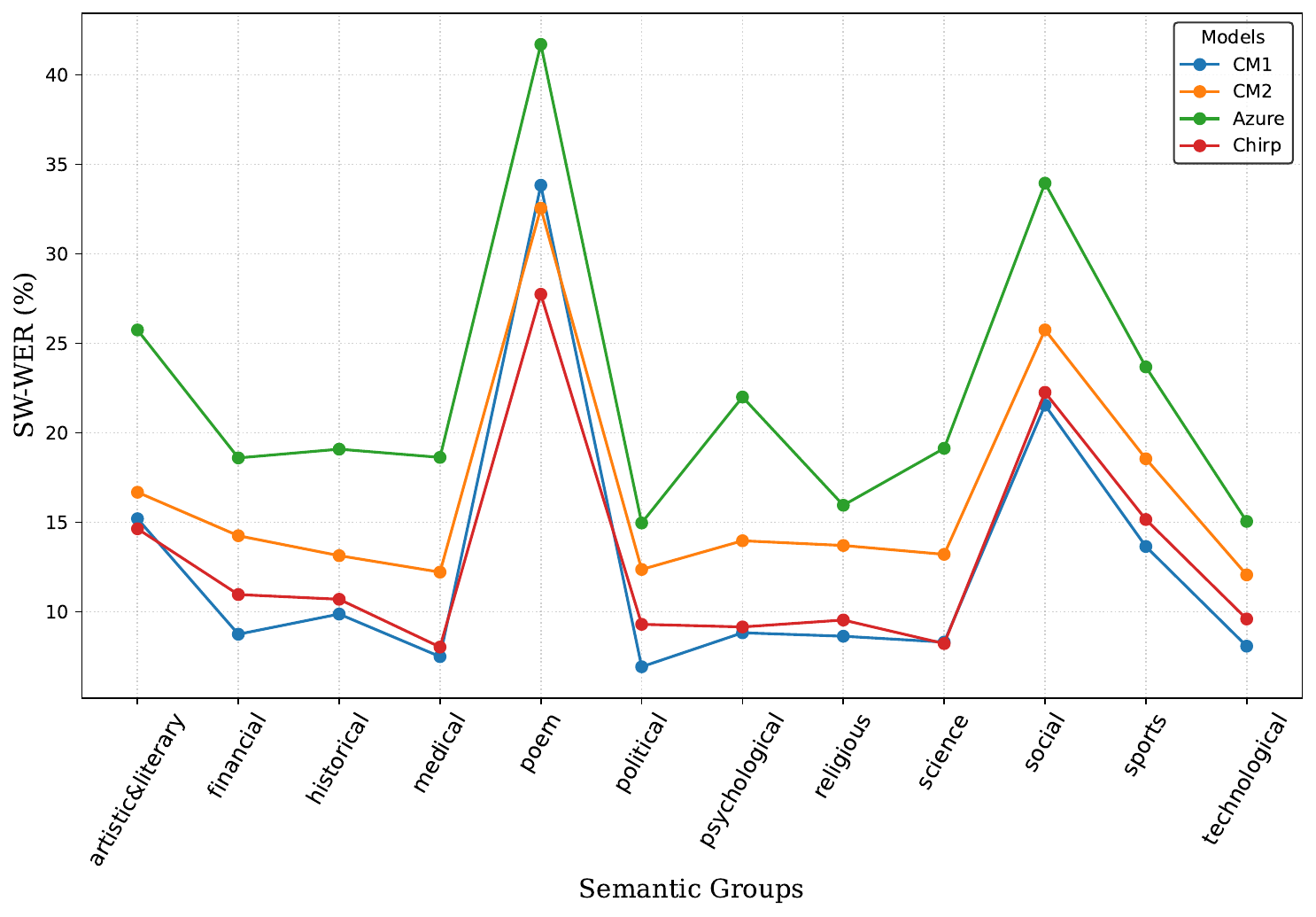}
        \caption{Commercial models}
    \end{subfigure}
    \begin{subfigure}[b]{0.48\textwidth}
        \centering
        \includegraphics[width=\linewidth]{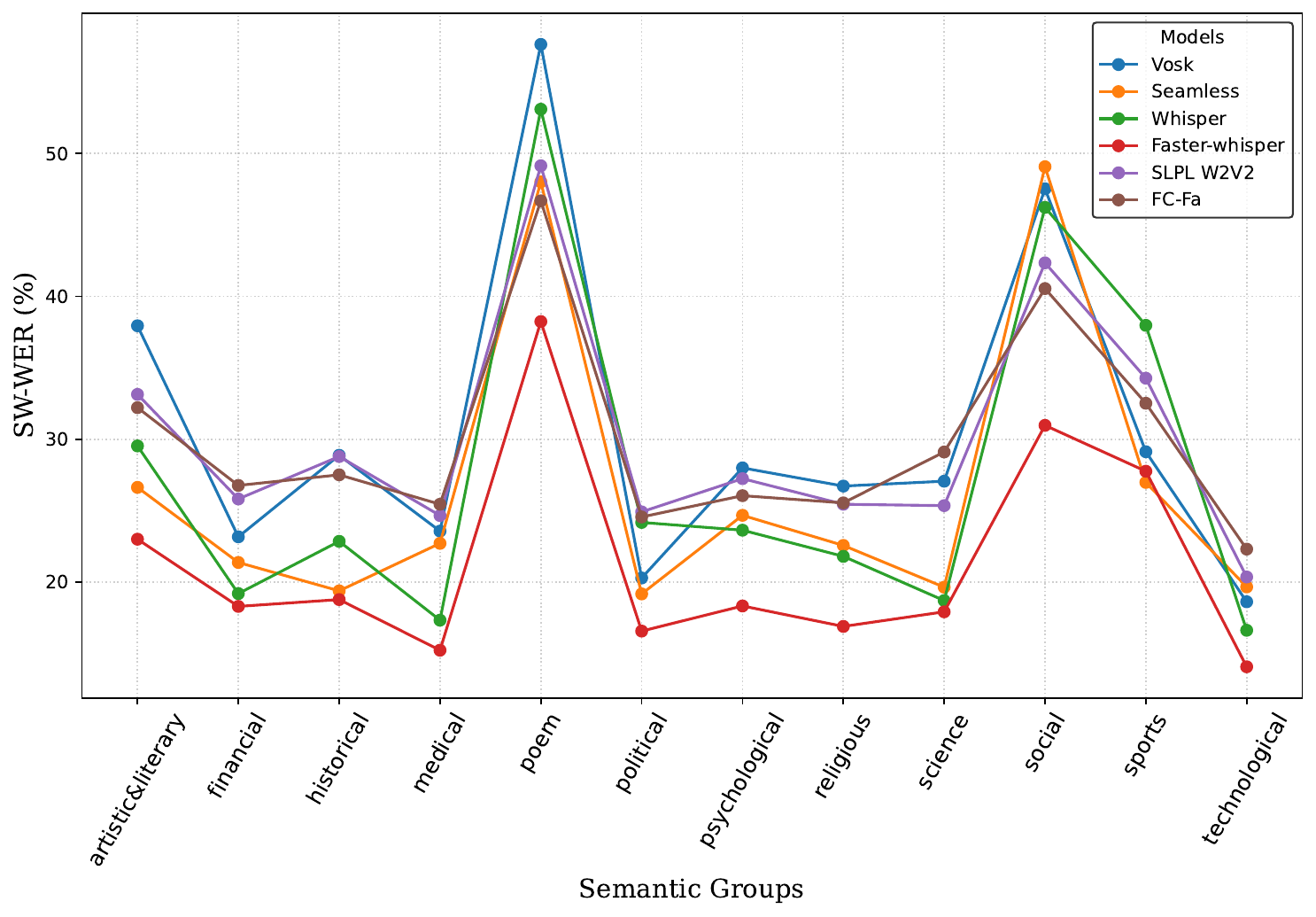}
        \caption{Open-source models}
    \end{subfigure}
    \caption{ASR models’ SW-WER across semantic domains (e.g., poem, social, technological), indicating higher errors in poem and social domains versus technological.}
    \label{fig:semanticcontent}
    \vspace{-3mm}
\end{figure}

The results in Figure \ref{fig:semanticcontent} show the performance of the models in different domains. In most models, the SW-WER error peaks in the “poem” and “social” domains, suggesting that these domains face particular challenges for most models. This can be attributed to the irregular structures and ambiguous phrasing often present in poetic and social expressions. In contrast, domains such as "technological" and "historical" exhibit relatively lower SW-WER for many models, possibly indicating a closer alignment between their linguistic patterns and the training data of these ASR systems. Additionally, certain models, such as whisper, demonstrate consistently higher error rates across multiple domains, whereas models like Avanegar and Aipa show lower SW-WER, suggesting stronger generalization or domain adaptation capabilities.

\section{Discussion and Conclusion}\label{sec7}

This study introduced PSRB, a comprehensive benchmark designed to enhance the evaluation and development of Persian ASR systems. Our analysis of multiple ASR models revealed systemic biases and performance limitations across demographic and linguistic factors, underscoring the need for more diverse and representative training data. Key challenges identified include word boundary errors, gender- and age-related biases, and domain-specific performance gaps, all of which provide valuable insights for improving ASR robustness and generalization.

Error analysis played a crucial role in identifying linguistic challenges specific to Persian ASR, such as difficulties in recognizing word boundaries, He-Kasreh errors, mismatches in formality, and instances of hallucination. These findings suggest that current ASR models lack sufficient linguistic normalization and context-aware processing, both of which are essential for accurate transcription in Persian. To address these issues, we introduced a new metric, SW-WER, which is robust to these errors and provides a more precise evaluation of model performance by weighting substitutions based on character-level differences. Furthermore, some models, particularly multilingual ones such as Seamless, exhibited hallucination errors, producing incorrect transcriptions or generating words not present in the audio. This raises concerns about their reliability in domains that require precise textual accuracy, such as legal and medical applications.

Regarding demographic factors, while most models demonstrated robustness to gender differences, all struggled with children's speech due to the lack of sufficient and balanced training data. Additionally, models performed well on standard Persian but exhibited weaker performance with regional accents such as Kurdish, Turkish, and Baluchi. The influence of speech spontaneity and data sources was also evident, with structured speech from audiobooks and news yielding lower error rates, whereas spontaneous speech from talk shows and films resulted in higher error rates. Moreover, the integration of a language model improved transcription accuracy by refining word predictions and sentence structures, highlighting the importance of further adapting models to Persian’s linguistic nuances.

Future research should investigate the effects of fine-tuning and explore the impact of combining different training datasets on model robustness. Traditional methods of data collection for ASR training primarily focus on covering a wide range of acoustic environments. However, for an ASR system to achieve high precision and performance, its training data must also exhibit semantic diversity. Enhancing both acoustic and semantic diversity in training datasets will contribute to the development of more accurate, inclusive, and resilient ASR technologies for Persian and other low-resource languages.

\bibliography{main}%
\end{document}